\documentclass[12pt,showpacs,a4paper,english]{revtex4}
\usepackage[T1]{fontenc}
\usepackage[latin1]{inputenc}
\usepackage{babel}
\usepackage{graphics}
\usepackage{setspace}

\begin{document}

\title{Pion production in deeply virtual Compton scattering}

\author{P.A.M. Guichon}
\affiliation{SPhN/DAPNIA, CEA Saclay, F91191 Gif-sur-Yvette, France}
\author{L. Moss\'e}
\affiliation{SPhN/DAPNIA, CEA Saclay, F91191 Gif-sur-Yvette, France}
\author{M. Vanderhaeghen}
\affiliation{Institut f\"ur Kernphysik, 
Johannes Gutenberg Universit\"at, D-55099 Mainz, Germany}

\date{\today}

\begin{abstract}
Using a soft pion theorem based on chiral symmetry and a 
$\Delta(1232)$ resonance
model we propose an estimate for the production cross section of low
energy pions in the deeply virtual Compton scattering (DVCS) process. 
In particular, we express the $e p \to e \gamma \pi N$ processes in
terms of generalized parton distributions. We provide estimates of the
contamination of the $e p \to e \gamma p$ DVCS observables due to this
associated pion production processes when the experimental data are
not fully exclusive, for a set of kinematical conditions
representative of present or planned experiments at JLab, HERMES and
COMPASS.  
\end{abstract}
\maketitle

\section{Introduction\label{Intro}}

The past few years have witnessed an intense theoretical activity
in the field of Generalized Parton Distributions (GPDs). They parametrize
the non perturbative content of the hadrons and are essentially matrix
elements of the form \cite{Ji97b,Rad96a}

\begin{equation}
\label{Eq1.1}
\left\langle N'\left| \bar{q}_{\alpha }(x^{\mu })q_{\beta }(0)\right| N\right\rangle ,
\end{equation}
where \( x^{\mu } \) is a light-like vector (\( x^{2}=0) \). In
(\ref{Eq1.1}) \( q_{\alpha } \) is the \texttt{}quark field with
Dirac index \( \alpha  \) and \( N,\, N' \) are two hadronic states
which can differ either by their structure or by the kinematics. The
limit \( N=N' \) gives the ordinary parton distributions.

In the Bjorken limit the GPDs appear in the amplitudes of exclusive
reactions such as\begin{equation}
\label{Eq1.2}
\gamma ^{*}(q)+N\rightarrow M+N',
\end{equation}
 where \( \gamma ^{*} \) is a highly virtual photon (\( Q^{2}=-q_{\mu }q^{\mu }\rightarrow \infty  \)
) and \( M \) can be a meson or a photon, in which case the reaction
is called Deeply Virtual Compton Scattering (DVCS). It is widely believed
that DVCS is conceptually the cleanest process to access the GPDs.
The basic reason is that the amplitude for meson production involves
the not so well known meson wave function. By contrast the final real
photon in DVCS can be considered as pointlike because photon structure
effects (VDM-like) are suppressed by powers of \( 1/Q^{2} \) \cite{Ji98a,Rad98,Col99} .

This theoretical simplicity of DVCS is to some extent counterbalanced
by its experimental difficulty which is certainly greater than in
the case of meson production. Nevertheless a number of experimental
attempts to measure photon electroproduction off the proton have been
performed or are in progress~\cite{Air02,Step02,Halla,Compass,H1zeus}. 
For the interpretation of these experiments
to be fruitful it is compulsory to have a control on the exclusivity
of the final state. If we consider DVCS on the proton, the most interesting
final state is the proton itself,
\begin{equation}
\label{Eq1.3}
\gamma ^{*}+p\rightarrow \gamma +p
\end{equation}
 and from now on we restrict our attention to this case. For further
reference we call it {}``Elastic DVCS''.
\footnote{Strictly speaking, exclusivity then means that there is one photon
and one proton in the final state and nothing else. In practice the
final state always contains soft photons which build the radiative
tail but their effect is included in the calculable radiative corrections.
So we ignore this subtility here and take Eq.(\ref{Eq1.3}) \emph{stricto
sensu} for what we mean by exclusive.
}

The problem for most experiments is that the experimental energy resolution is not good
enough to isolate the exclusive \( (\gamma +p) \) channel because
it is separated from the first strong inelastic channel by only the
small pion mass. In practice such data will always be contaminated
by the reactions, referred to as Associated DVCS (ADVCS)~: 
\begin{equation}
\label{Eq1.4}
\gamma ^{*}+p\rightarrow \gamma +p+\pi ^{0}\, \, \, {\rm or}\, \, \gamma ^{*}+p\rightarrow \gamma +n+\pi ^{+},
\end{equation}
where \( \pi ^{+,0} \) is a low energy pion which escapes detection.
Note that experiments which have sufficient energy resolution
to distinguish the $\gamma \pi N$ final state from the $\gamma p$ one 
can study the process (\ref{Eq1.4}) for itself \cite{Michel03}, thereby
enlarging the scope of virtual Compton scattering. 
This is likely to be the case of the experiments planned at Jlab.

In this paper we propose a calculation of the cross section for reaction
(\ref{Eq1.4}) and we compare it with the elastic channel (\ref{Eq1.3})
by integrating over the pion momentum up to a given cutoff. As the dangerous
pions are those which have a small energy we use the soft pion techniques
based on chiral symmetry. In the soft pion limit, that is \( k_{\pi }\rightarrow 0 \)
where \( k_{\pi } \) is the pion 4-momentum, this allows to evaluate
the associated DVCS amplitude using the \emph{same} generalized parton
distribution as in the elastic case. So, in a relative sense, this
evaluation is model independent. The inherent limitation of this approach
is that, apart the chiral limit (\( m_{\pi }\rightarrow 0 \) ), it
gives a reliable estimate only for a small center of mass (CM) energy
of the final pion-nucleon pair. Typically the upper limit is set by
the excitation energy of the first \( \Delta  \) resonance, that
is about 300 MeV. In order to increase the range of validity of our estimate
we propose a (model dependent) estimate of the associated DVCS corresponding
to\[
\gamma *+p\rightarrow \gamma +\Delta .\]
For this we use the large \( N_{c} \) limit as a guidance which allows
one to relate the GPDs of the \( N\rightarrow \Delta  \) transition
to the ones of the \( N\rightarrow N \) transition \cite{Fra00}.

In practice one measures the \( (l,\, l',\, \gamma ) \) reaction
where \( l \) denotes either a muon or an electron. The amplitude
\( T^{ll'\gamma } \) for this reaction is the coherent sum of \( T^{VCS} \),
the virtual Compton scattering amplitude and of \( T^{BH} \) the
amplitude of the Bethe-Heitler process where the final photon is emitted
by the lepton. When the reaction produces only a proton (elastic case)
the calculation of this {}``elastic-BH'' amplitude only involves
the elastic form factors of the proton, which are well known. When
a pion is produced together with the final nucleon, then the corresponding
{}``associated-BH'' (ABH) amplitude involves the pion electro-production
amplitude. For consistency reasons we shall evaluate this amplitude
in the same framework as the associated DVCS.

Our paper is organized as follows: in Section \ref{Preliminary} we
specify the kinematics and give the relevant expressions for the amplitudes
and cross sections. In Section \ref{Handbag} we remind the leading
order approximation for the DVCS amplitude and define the twist-2
quark operators which control the DVCS amplitude both in the elastic
and associated case. In Section \ref{GPD} we remind the definition
of the GPDs in the elastic case. In Section \ref{LET} we derive the
soft pion theorems relevant for the evaluation of the associated DVCS
and BH amplitudes. In Section \ref{ADVCS_Delta} we present a model
to estimate the associated DVCS and BH in the \( \Delta  \) region.
Section \ref{Results} is devoted to a presentation and discussion
of our results in kinematical situations of interest. Section \ref{Conclusion}
is our conclusion.

\section{Preliminary\label{Preliminary}}

\begin{table}

\caption{\label{table_1}External particles variables.}

\begin{tabular}{|c|c|c|c|c|c|c|}
\hline 
&
Initial &
Final &
Final&
Initial&
Final&
Final\\
&
lepton&
lepton&
photon&
proton&
nucleon&
pion\\
\hline
\hline 
Momentum&
\( k \)&
\( k' \)&
\( q' \)&
\( p \)&
\( p' \)&
\( k_{\pi } \)\\
\hline 
Mass&
\( m_{l} \)&
\( m_{l} \)&
\( 0 \)&
\( M \)&
\( M \)&
\( m_{\pi } \)\\
\hline 
Helicity, spin&
\( h \)&
\( h' \)&
\( \lambda ' \)&
\( \sigma  \)&
\( \sigma ' \)&
\\
\hline 
Wave function&
\( u(k,h) \)&
\( u(k',h') \)&
\multicolumn{1}{c|}{\( \varepsilon '(q',\lambda ') \)}&
\( u(p,\sigma ) \)&
\( u(p',\sigma ') \)&
\\
\hline
\end{tabular}
\end{table}

In Table \ref{table_1} we have collected the characteristics of the
particles involved in the reaction. We define \( q \) as the momentum
of the virtual photon exchanged in the VCS process, that is \( q=k-k' \).
So the BH virtual photon has \( q-q' \). The relevant Lorentz scalars
are\begin{equation}
\label{Eq_prelim.5}
Q^{2}=-q^{2},\, \, \, t_{\gamma }=(q-q')^{2},\, \, \, x_{B}=\frac{Q^{2}}{2p.q},\, \, \, W^{2}=(p+q-q')^{2}.
\end{equation}
We shall also note \( t=(p'-p)^{2} \) which coincides with \( t_{\gamma } \)
in the elastic case or in the limit \( k_{\pi }=0. \)
\begin{figure}
{\centering \resizebox*{0.6\textwidth}{!}{\includegraphics{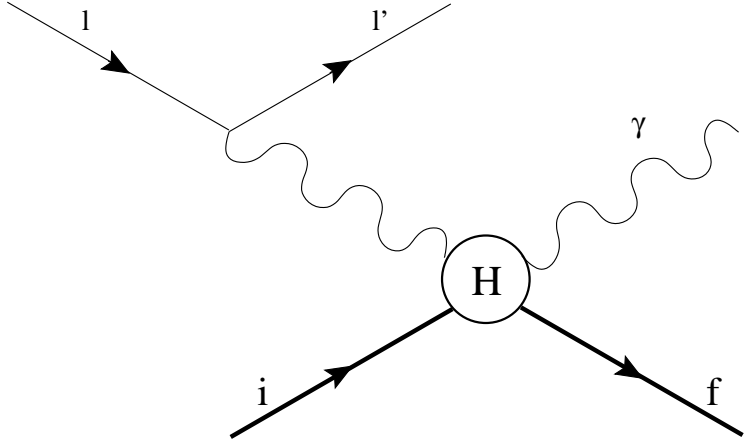}} \par}

\caption{\label{Fig_VCS}Graph of the virtual Compton process}
\end{figure}

\begin{figure}
{\centering \resizebox*{1\textwidth}{!}{\includegraphics{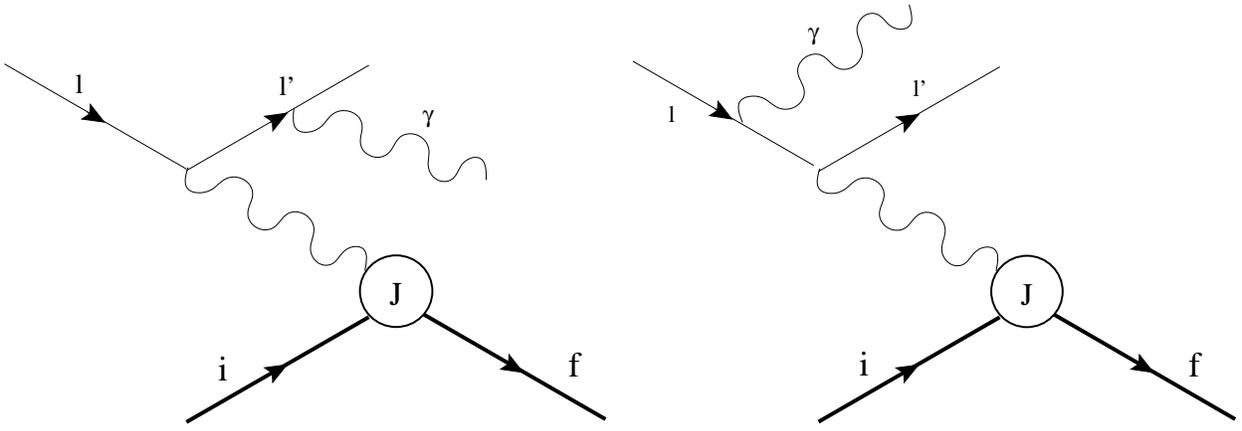}} \par}

\caption{\label{Fig_BH}Graphs of the Bethe-Heitler process}
\end{figure}
In the one photon exchange approximation and in the Lorentz gauge
we have the generic expressions for the VCS and BH amplitudes~:\begin{eqnarray}
T^{VCS} & = & \pm e^{3}\varepsilon ^{\prime *}_{\mu }H^{\mu \nu }\frac{1}{Q^{2}}\bar{u}(l')\gamma _{\nu }u(l),\label{Eq_prelim.1.1} \\
T^{BH} & = & -e^{3}\, J_{\nu }\, \frac{1}{t_{\gamma }}\, \bar{u}(l')\left[ \gamma .\varepsilon ^{\prime *}\frac{1}{\gamma .(k'+q')-m_{l}}\gamma ^{\nu }+\right. \nonumber \\
 &  & \, \, \, \, \, \, \, \, \, \, \, \, \, \, \, \, \, \, \, \, \, \, \, \, \, \, \, \, \, \, \, \, \, \, \, \, \, \left. \gamma ^{\nu }\frac{1}{\gamma .(k-q')-m_{l}}\gamma .\varepsilon ^{\prime *}\right] u(l)\label{Eq_prelim.1.2} \\
T^{ll'\gamma } & = & T^{VCS}+T^{BH}\label{Eq_prelim.1.3} 
\end{eqnarray}
They correspond to the graphs of Fig.\ref{Fig_VCS} and \ref{Fig_BH}
respectively. In the above expressions the charge \( e=\sqrt{4\pi /137} \)
has been factored out and the \( \pm  \) sign refers to the charge
of the lepton beam. The spinors are normalized to \( \bar{u}u=2m \)
with \( m \) a generic mass and the photon polarisation is normalized
to \( \varepsilon ^{\prime}.\varepsilon ^{\prime}=-1 \). The
amplitudes \( H \) and \( J \) are defined by \begin{eqnarray}
H^{\mu \nu } & = & -i<f|\int d^{4}xe^{-iq.x}T\left[ j^{\mu }(0),j^{\nu }(x)\right] |i>,\label{Eq_prelim.2.1} \\
J^{\mu } & = & <f|j^{\mu }(0)|i>,\label{Eq_prelim.2.2} 
\end{eqnarray}
where \( i,\, f \) are the appropriate hadronic states and \( j(x) \)
is the electromagnetic current carried by the quarks\begin{equation}
\label{Eq_prelim.3}
j^{\mu }(x)=\bar{q}\gamma ^{\mu }{\cal Q}q(x)=\sum _{f}{\cal Q}_{f}\bar{q}_{f}\gamma ^{\mu }q_{f}(x),
\end{equation}
 with \( \cal Q \) the diagonal charge matrix \( {\cal Q}=[2/3,-1/3,\cdots ] \).
The matrix element of the electromagnetic current between nucleon
states has the usual form factor decomposition~:\begin{equation}
\label{Eq_prelim.3.1}
<N(p')|j^{\mu }(0)|N(p)>=\bar{u}(p')\left[ F_{1}^{p,n}(t)\gamma ^{\mu }+iF_{2}^{p,n}(t)\frac{\sigma ^{\mu \nu }(p'-p)_{\nu }}{2M}\right] u(p)
\end{equation}
where \( p,n \) refer to the proton or the neutron and \( t=(p-p')^{2} \).
With our normalizations we have \( F_{1}^{p}(0)=1,\, F_{1}^{n}(0)=0,\, F_{2}^{p}(0)=1.79,\, F_{2}^{n}(0)=-1.91 \). 

We shall note \( T^{ll'\gamma }_{el.} \) the amplitude for a final
hadronic state containing only a proton and \( T^{ll'\gamma }_{as.}(\pi ^{+,0}) \)
the amplitude for producing a nucleon and a pion of charge +1 or 0.
A similar notation will be used for the cross sections. 

To fix the idea, we consider an experiment where \( (k',q') \) the
momenta of the final lepton and photon are measured while the final
hadronic state \( p' \) or \( (p',k_{\pi }) \) is not observed.
The elastic \( (ll'\gamma ) \) event is selected by the missing
mass condition\begin{equation}
\label{Eq_prelim.4}
W^{2}=(k-k'+p-q')^{2}=M^{2}.
\end{equation}
 In pratice the events will be integrated up to a cutoff \( W_{max} \)
which generally exceeds the pion production threshold. In the following
we assume that the experimental resolution on $W$ is nevertheless
good enough so that the production of more than one pion can be neglected.

If we assume that the one particle states are normalized as~:\begin{equation}
\label{Eq_prelim.5.1}
<k|k'>=(2\pi )^{3}2k_{0}\delta (\vec{k}-\vec{k}'),
\end{equation}
we have the following expressions for the invariant cross sections~:\begin{eqnarray}
d\sigma ^{ll'\gamma }_{el.} & = & d^{4}\Gamma \delta (W-M)dW\left| T^{ll'\gamma }_{el.}\right| ^{2},\label{Eq_prelim.6} \\
d\sigma ^{ll'\gamma }_{as.}(\pi ^{+,0}) & = & d^{4}\Gamma dW\frac{\left| \vec{k}^{*}_{\pi }\right| }{16\pi ^{3}}\int d\hat{k}^{*}_{\pi }\left| T^{ll'\gamma }_{as.}(\pi ^{+,0})\right| ^{2}.\label{Eq_prelim.7} 
\end{eqnarray}
In the above equations the common phase space factor%
\footnote{This expression is for a full azimuthal coverage for the lepton detection.
For a finite azimuthal coverage \( \Delta \phi  \), it should be
multiplied by \( \Delta \phi /2\pi . \)
} is\begin{equation}
\label{Eq_prelim.8}
d^{4}\Gamma =\frac{dx_{B}dQ^{2}dt_{\gamma }d\Phi }{128(2\pi )^{4}(p.k)^{2}x_{B}\sqrt{1+4M^{2}x_{B}^{2}/Q^{2}}}
\end{equation}
where \( \Phi \) is the angle between the planes \( (\vec{k},\vec{k}\, ') \)
and \( (\vec{q},\vec{q}\, ') \) and \( \vec{k}^{*}_{\pi }=\left| \vec{k}^{*}_{\pi }\right| \hat{k}^{*}_{\pi } \)
stands for the pion momentum in the frame defined by \( \vec{p}+\vec{q}-\vec{q}\, '=0, \)
that is the rest frame of the final pion-nucleon pair. One has \begin{equation}
\label{Eq_prelim.9}
\left| \vec{k}^{*}_{\pi }\right| ^{2}=\frac{W^{4}-2W^{2}(M^{2}+m_{\pi }^{2})+(M^{2}-m_{\pi }^{2})^{2}}{4W^{2}}.
\end{equation}

From Eqs.~(\ref{Eq_prelim.6},\ref{Eq_prelim.7}) we see that the contamination
of the elastic reaction by the associated one is measured by the dimensionless
quantity 
\begin{equation}
\label{Eq_prelim.10}
\kappa (Q^{2},x_{B},t_{\gamma },\Phi)=\frac{1}{16\pi ^{3}}\frac{\int _{M+m_{\pi }}^{W_{max}}dW\left| \vec{k}^{*}_{\pi }\right| \int d\hat{k}^{*}_{\pi }\sum _{hel.}\left( \left| T^{ll'\gamma }_{as.}(\pi ^{+})\right| ^{2}+\left| T^{ll'\gamma }_{as.}(\pi ^{0})\right| ^{2}\right) }{\sum _{hel}\left| T^{ll'\gamma }_{el.}\right| ^{2}}.
\end{equation}
 where \( \sum _{hel} \) denotes the sum over all the helicities.
This is the quantity we want to evaluate.

\section{Handbag approximation for the VCS amplitude\label{Handbag}}

\begin{figure}
{\centering \resizebox*{0.7\textwidth}{!}{\includegraphics{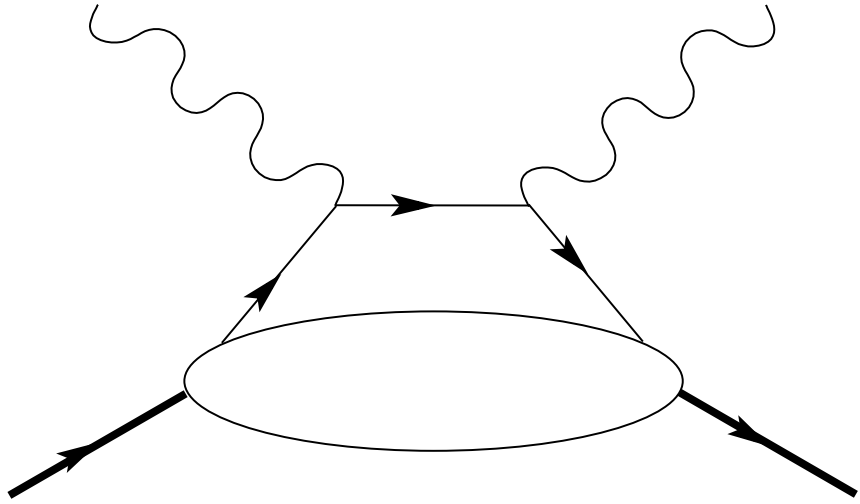}} \par}
\caption{\label{Fig_handbag}The handbag approximation to the VCS amplitude.
The crossed term is not shown.}
\end{figure}

We need to evaluate the hadronic tensor \( H^{\mu \nu } \) (Eq.\ref{Eq_prelim.2.1})
in the generalized Bjorken limit which here is defined by~: \begin{equation}
\label{Eq_leading.0}
Q^{2}\rightarrow \infty ,\, x_{B}\, {\rm fixed},\, t_{\gamma }/Q^{2}\rightarrow 0,\, W^{2}/Q^{2}\rightarrow 0,
\end{equation}
 where the two last conditions are necessary to have factorisation
of the amplitude \cite{Col99}. The condition that \( W \) remains
small with respect to \( Q \) is not explicitely stated for elastic
VCS since \( W=M \) in this case, but it is necessary for the associated
VCS. The factorisation theorem garantees that the amplitude factorizes
in a hard part which can be computed in perturbation theory as a series
in \( \alpha _{S}(Q^{2}) \) and a soft part which depends on the
long distance structure of the hadron. In this work we consider only
the leading term of the hard part which amounts to evaluate the amplitude
in the handbag approximation as shown in Fig. \ref{Fig_handbag}

For this purpose we adapt the formulation of \cite{Ji97b} to our
problem. In the Bjorken limit both the elastic and associated VCS
are light-cone dominated. Therefore it is convenient to introduce
two light-like vectors \( \tilde{n} \) and \( \tilde{p} \) to record
the flow of hard momentum. These Sudakov vectors are chosen to be
in the hyper-plane defined by the virtual photon momentum \( q \)
and another vector \( P \) which is related to the target-ejectile
motion. In our case it is convenient to choose~:
\begin{eqnarray}
P & = & \frac{p+p'}{2}\, {\rm for\, \, the\, \, elastic\, \, VCS},
\label{Eq_leading.0.1} \\
P & = & \frac{p+p'+k_{\pi }}{2}\, {\rm for\, \, the\, \, associated\,\, VCS},
\label{Eq_leading.0.2} 
\end{eqnarray}
so that, for both reactions, \( P \) has the same expression in
terms of \( p,\, q,\, q' \) that is~: 
\begin{equation}
\label{Eq_leading.1}
P=p+\frac{q-q'}{2}=p+\frac{\Delta }{2}.
\end{equation}
 Other choices of \( P \) are allowed, for instance \( P=p \), but
the formulation is more symmetric if one uses the definition (\ref{Eq_leading.1}).
One imposes the normalization conditions~:
\footnote{The normalization \( \tilde{n}.P =1\) is in fact irrelevant. The choice
\( \tilde{n}.P \)=1 is a matter of convenience.
}\begin{equation}
\label{Eq_leading.2}
\tilde{n}^{2}=\tilde{p}^{2}=0,\, \, \, \tilde{n}.\tilde{p}=1,\, \, \,
\tilde{n}.P=1,
\end{equation}
 and one defines the variable \( \xi ' \) as~:
\begin{equation}
\label{Eq_leading.2.1}
\xi '=\frac{Q^{2}}{4P.q}=\frac{P.q}{2P^{2}}\left[ -1+\sqrt{1+\frac{Q^{2}P^{2}}{\left( P.q\right) ^{2}}}\right] \approx \frac{x_{B}/2}{1-x_{B}/2}{\rm \, \, in\, \, the\, \, Bjorken\, \, limit.}
\end{equation}
 From this one gets the decomposition~:
\begin{eqnarray}
P & = & \tilde{p}+\frac{P^{2}}{2}\tilde{n},\nonumber \\
q & = & -2\xi '\, \tilde{p}+\frac{Q^{2}}{4\xi '}\tilde{n},\label{Eq_leading.4} 
\end{eqnarray}
where \begin{equation}
\label{Eq_leading.5}
P^{2}=P^{\mu }P_{\mu }=\frac{M^{2}+W^{2}}{2}-\frac{t_{\gamma }}{4}.
\end{equation}
An arbitrary vector \( a \) has the covariant decomposition~: \begin{equation}
\label{Eq_leading.3}
a=a.\tilde{n}\, \tilde{p}+a.\tilde{p}\, \tilde{n}+a(\bot ),
\end{equation}
with the \( \bot  \) component defined by \( a(\bot ).\tilde{n}=a(\bot ).\tilde{p}=0 \).
If one restricts to the case where the final photon is real one then
gets~:

\[
\Delta =-2\xi \tilde{p}+\left( \xi P^{2}+\frac{W^{2}-M^{2}}{2}\right) \tilde{n}+\Delta (\bot ),\]
 where~: \[
\xi =\xi '\frac{Q^{2}-t_{\gamma }-2\xi '\, (W^{2}-M^{2})}{Q^{2}+4\xi '^{2}P^{2}}\approx \xi '{\rm \, \, in\, \, the\, \, Bjorken\, \, limit.}\]

If one keeps the leading light-cone singularity of the time ordered
product of currents in Eq. (\ref{Eq_prelim.2.1}) one has, at leading
order in \( \alpha _{s}(Q^{2}) \) \cite{Ji97a}~:\begin{eqnarray}
 &  & H^{\mu \nu }=\int dx\, C_{+}^{\mu \nu }(x,\xi )<f|\sum _{f}{\cal Q}_{f}^{2}S_{f}(x,\tilde{n})|i>\nonumber \\
 &  & +\int dx\, C_{-}^{\mu \nu }(x,\xi )<f|\sum _{f}{\cal Q}_{f}^{2}S^{5}_{f}(x,\tilde{n})|i>+O\left( \frac{1}{Q}\right) ,\label{Eq_leading.6} 
\end{eqnarray}
 where the hard scattering coefficients (in which we consistently make the
 approximation $\xi \approx \xi'$) write~: 
\begin{eqnarray}
C_{+}^{\mu \nu }(x,\xi ) & = & \frac{1}{2}(\tilde{n}^{\mu
}\tilde{p}^{\nu }+\tilde{n}^{\nu }\tilde{p}^{\mu }-g^{\mu \nu })\left(
\frac{1}{x- \xi +i\varepsilon }
+\frac{1}{x + \xi -i\varepsilon }\right) ,\label{Eq_leading.7.1} \\
C_{-}^{\mu \nu }(x,\xi ) & = & 
\frac{i}{2}\varepsilon ^{\mu \nu \rho \sigma }
\tilde{p}_{\rho }\tilde{n}_{\sigma }\left( \frac{1}{x-\xi +i\varepsilon }
-\frac{1}{x+ \xi -i\varepsilon }\right) ,\label{Eq_leading.7.2} 
\end{eqnarray}
The twist-2 operators \( S_{f} \) and \( S_{f}^{5} \), for which
we often use the global notation \( S^{(5)}_{f} \) or \( S^{(5)} \),
are defined by~: \begin{eqnarray}
S_{f}(x,\tilde{N}) & = & \int \frac{d\lambda }{2\pi }e^{i\lambda x}\bar{q}_{f}(-\lambda \tilde{N}/2)\gamma .\tilde{N}\, q_{f}(\lambda \tilde{N}/2),\label{Eq_leading.8.1} \\
S_{f}^{5}(x,\tilde{N}) & = & \int \frac{d\lambda }{2\pi }e^{i\lambda x}\bar{q}_{f}(-\lambda \tilde{N}/2)\gamma .\tilde{N}\, \gamma ^{5}q_{f}(\lambda \tilde{N}/2)\label{Eq_leading.8.2} 
\end{eqnarray}
 with \( \tilde{N} \) an arbitrary light-like vector. They obviously
satisfy the scaling law~:\begin{equation}
\label{Eq_leading.8.3}
S^{(5)}(\omega x,\omega \tilde{N})=S^{(5)}(x,\tilde{N}).
\end{equation}
 In Eq.(\ref{Eq_leading.6}) one has \( \tilde{N}=\tilde{n} \), which
is the only relic of the hard scattering in the soft matrix elements.

\begin{figure}
{\centering \resizebox*{0.7\textwidth}{!}{\includegraphics{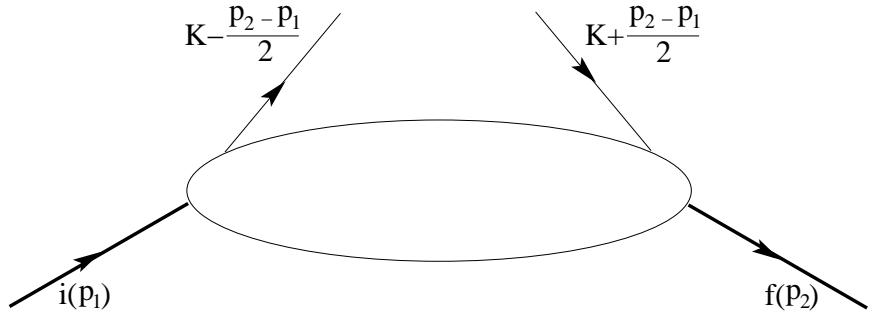}} \par}
\caption{\label{Fig_twist2}Representation of the twist-2 operator.}
\end{figure}
In Fig.~\ref{Fig_twist2} we have represented the matrix element
of \( S^{(5)} \) between generic hadronic states \( i(p_{1}) \)
and \( f(p_{2}) \). The quark lines are on mass shell Fock states
quantized at equal light cone time \( x_{\mu }\tilde{N}^{\mu }=0. \)
If one labels the momenta of the initial and final active quarks as
\( K-(p_{2}-p_{1})/2 \) and \( K+(p_{2}-p_{1})/2 \) respectively,  
then the integration over \( \lambda  \) in 
Eqs.~(\ref{Eq_leading.8.1},\ref{Eq_leading.8.2})
implies that \( x=K.\tilde{N} \). On the other hand, from the additivity
of the longitudinal momentum, one has~: \[
-\frac{\left| (p_{1}+p_{2}).\tilde{N}\right| }{2}\leq K.\tilde{N}\leq \frac{\left| (p_{1}+p_{2}).\tilde{N}\right| }{2},\]
So, if one chooses the normalization \( \tilde{N}.(p_{1}+p_{2})/2=1 \)
as in (\ref{Eq_leading.2}), the integration over \( x \) is effectively
restricted to the interval \( [-1,1] \) in Eq.~(\ref{Eq_leading.6}).

In Eq.~(\ref{Eq_leading.6}) the initial state is always the nucleon,
with momentum \( P-\frac{\Delta }{2}=p \). On the other hand the
final state, which has momentum \( P+\frac{\Delta }{2}=p+\Delta  \),
can be either the nucleon (\( p+\Delta =p' \)) or the pion-nucleon
system (\( p+\Delta =p'+k_{\pi } \)). Note that for given \( p,\, q,\, q' \)
the Sudakov vectors are the same in both cases. This is not essential
but it simplifies the formulation.

Finally, we point out that because expression (\ref{Eq_leading.6})
is accurate up to terms of order \( 1/Q \) gauge invariance is not
strictly satisfied. One finds instead~:\[
q'_{\mu }H^{\mu \nu }\sim \Delta (\perp ),\, \, \, H^{\mu \nu }q_{\nu }=0.\]
However it has been shown in \cite{Vdh99} that 
electromagnetic gauge invariance is
restored by adding to \( H^{\mu \nu } \) a term which is explicitely
of order \( \sqrt{- t_{\gamma }}/Q \), and therefore does not contribute in
the Bjorken limit. The effect of this extra term has been calculated
for realistic situations and found to be negligible\cite{Vdh99}.
So gauge invariance is not a serious issue as long as the conditions
(\ref{Eq_leading.0}) are satisfied.

\section{Generalized partons distributions for the elastic case\label{GPD}}

The matrix elements of \( S^{(5)} \) are the non perturbative inputs
of the calculation. If we note \( N(p_{1}),N(p_{2}) \) two generic
nucleon states, the elastic matrix elements are parametrized according
to~\cite{Ji97b}~:\begin{eqnarray}
 &  & <N(p_{2})|S_{f}(x,\tilde{N})|N(p_{1})>=\nonumber \\
 &  & \frac{1}{P_{12}.\tilde{N}}\bar{u}(p_{2})\left[ H_{f/N}\gamma .\tilde{N}+iE_{f/N}\frac{\sigma ^{\mu \nu }\tilde{N}_{\mu }(p_{2}-p_{1})_{\nu }}{2M}\right] u(p_{1}),\label{Eq_leading.9.1} \\
 &  & <N(p_{2})|S_{f}^{5}(\kappa ,\tilde{N})|N(p_{1})>=\nonumber \\
 &  & \frac{1}{P_{12}.\tilde{N}}\bar{u}(p_{2})\left[
\tilde{H}_{f/N}\gamma.\tilde{N}
+\tilde{E}_{f/N}\frac{\tilde{N}.(p_{2}-p_{1})}{2M}\right]\gamma ^{5}u(p_{1}), 
\label{Eq_leading.9.2} 
\end{eqnarray}
where we note \( P_{12}=(p_{1}+p_{2})/2. \) The Generalized Parton
Distributions (GPDs) \( H,\, E,\, \tilde{H},\, \tilde{E} \) are
a priori functions of \( x \) and of the invariants which can be
formed with \( p_{1},\, p_{2} \) and \( \tilde{N} \), that is~: \[
GPD=GPD\left\{ x,\, (p_{2}-p_{1}).\tilde{N},\, P_{12}.\tilde{N},\, t_{12}=(p_{1}-p_{2})^{2}\right\} .\]
 However from the scaling law (\ref{Eq_leading.8.3}) and the definitions
(\ref{Eq_leading.9.1},\ref{Eq_leading.9.2}) we have~:\[
GPD\left\{ \omega x,\, \omega (p_{2}-p_{1}).\tilde{N},\, \omega P_{12}.\tilde{N},\, t_{12}\right\} =GPD\left\{ x,\, (p_{2}-p_{1}).\tilde{N},\, P_{12}.\tilde{N},\, t_{12}\right\} \]
so the GPDs are only function of \( t_{12} \) and of the ratios \( x/P_{12}.\tilde{N} \)
and \( (p_{2}-p_{1}).\tilde{N}/P_{12}.\tilde{N} \) . The standard
choice for the functional dependence is~: 
\begin{eqnarray}
&&GPD\left\{ x,\, (p_{2}-p_{1}).\tilde{N},\, P_{12}.\tilde{N},\,
t_{12}\right\} \nonumber \\
&&= \, GPD\left\{ x_{12}=\frac{x}{P_{12}.\tilde{N}},\, \xi
_{12}=-\frac{(p_{2}-p_{1}).\tilde{N}}{2P_{12}.\tilde{N}},\,
t_{12}\right\} \, .
\end{eqnarray}
 Of course in the elastic case one has \( x=x_{12} \) and \( \xi _{12}=\xi  \)
but when one of the nucleons is an intermediate state which is off
the energy shell, as is happens in the ADVCS process (see subsection
\ref{LET_twist2}), this is no longer true. 

The interest and properties of the GPDs are presented in several reviews
\cite{Ji98,Gui98,Rad01b,GPV01}. In Section \ref{Results} we explain the
model used for our estimates. Note that one can devise a parametrization
analogous to (\ref{Eq_leading.9.1},\ref{Eq_leading.9.2}) when the
final state is the \( \pi N \) system \cite{GPV01}, but for practical
purposes that is not very useful as it involves many new unknown functions.

For further use we introduce iso-vector GPDs through~:\begin{equation}
\label{Eq_leading.10}
S_{V,\alpha }(x,\tilde{N})=\int \frac{d\lambda }{2\pi }e^{i\lambda
x}\bar{q}(-\lambda \tilde{N}/2)\gamma .\tilde{N}\frac{\tau _{\alpha }}{2}\Pi
_{ud}q(\lambda \tilde{N}/2),
\end{equation}
 where \( (\tau _{\alpha },\, \alpha =1,2,3) \) are the Pauli matrices
and \( \Pi _{ud} \) project on the \( u,d \) part of the quark flavor
multiplet. The parametrisation in term of GPDs then writes\begin{equation}
\label{Eq_leading.11}
<N(p_{2})|S_{V,\alpha }(x,\tilde{N})|N(p_{1})>=\bar{u}(p_{2})\left[ H_{V}\gamma .\tilde{N}+\cdots \right] \frac{\tau _{\alpha }^{N}}{2}u(p_{1})
\end{equation}
where now the Dirac spinor is also a iso-spinor for the nucleon and
\( \tau ^{N} \) are the Pauli matrices acting on it. A similar definition
holds for \( S_{V}^{5} \). The fact that \( (H_{V},\, E_{V},\ldots ) \)
are independent of \( \alpha  \) and of the isospin of the initial
and final nucleons is a consequence of isospin symmetry. As a consequence
of their definitions, we have the following (generic) relations~:\begin{equation}
\label{Eq_leading.12}
H_{V}=H_{u/p}-H_{d/p}=H_{d/n}-H_{u/n}.
\end{equation}
The iso-vector GPDs are thus controlled by the valence quarks in so
far as the sea is flavor symmetric. The parametrisation defined by
Eqs.~(\ref{Eq_leading.10},\ref{Eq_leading.11}) is necessary to describe
reactions where the nucleon charge is changed, as is the case when
a charged pion is emitted.

\section{Soft pion theorems\label{LET}}

\subsection{Soft pion expansion}

To evaluate the amplitude for the reaction~:\[
l+N\rightarrow l'+\gamma +N+\pi , \]
we see from Eqs(\ref{Eq_prelim.1.1},\ref{Eq_prelim.1.2},\ref{Eq_prelim.2.2},\ref{Eq_leading.6})
that we need the matrix elements~:\begin{equation}
\label{Eq_LET.00}
<N\, \pi |j_{\mu }|N>,
\end{equation}
 which is the pion electro-production amplitude, and\begin{equation}
\label{Eq_LET.01}
<N\, \pi |\sum _{f}{\cal Q}_{f}^{2}S^{(5)}_{f}|N>,
\end{equation}
 which we call (loosely) the \emph{pion twist-2 production} amplitudes.
To simplify the notations we do not write the isospin label of the
pion or of the various iso-vector quantities if it is not necessary. 

\begin{figure}
{\centering \resizebox*{0.8\textwidth}{!}{\includegraphics{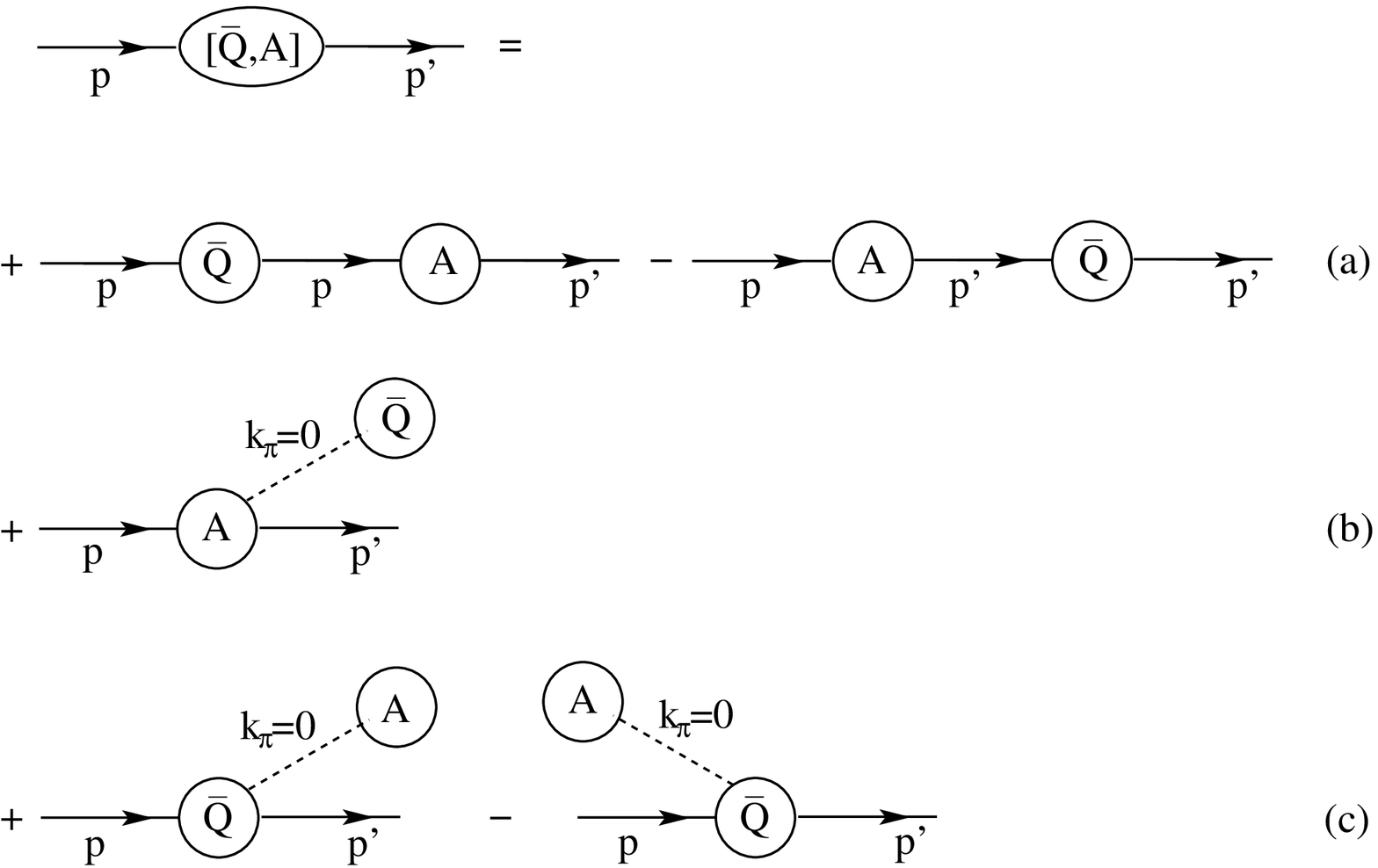}} \par}
\caption{\label{Fig_soft_pion} Soft pion expansion of the commutator. Note
that the intermediate lines are not propagators. }
\end{figure}

\begin{figure}
{\centering \resizebox*{0.7\textwidth}{!}{\includegraphics{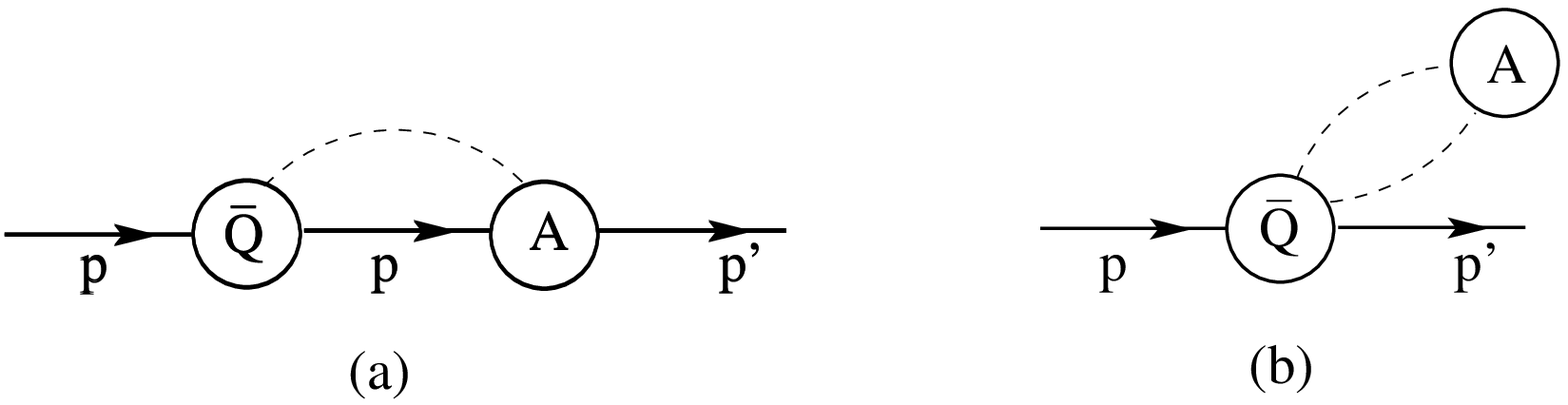}} \par}

\caption{\label{Fig_pion_loop}Example of pion loops. }
\end{figure}

We first recall the main steps in the derivation of a soft pion theorem
for a matrix element of the form~:\[
<N\, \pi (\vec{0})|A|N>,\]
where \( A \) is either the electromagnetic current, Eq.(\ref{Eq_LET.00}),
or one of the twist-2 operators, Eq.(\ref{Eq_LET.01}). In the approach
of de Alfaro \emph{et al.} \cite{dAlfaro68,Furlan72}, which we follow
here, one considers a physical pion at rest and looks for the dominant
contribution as \( m_{\pi }\rightarrow 0. \) In the chiral limit,
the isovector axial current~:%
\footnote{We remind that \( \Pi _{ud} \) simply projects on the \( u,d \)
part of the flavour multiplet
}\begin{equation}
\label{Eq_LET.1}
j_{\mu }^{5,\alpha }=\bar{q}\gamma _{\mu }\gamma _{5}\frac{\tau ^{\alpha }}{2}\Pi _{ud}q,\, \, \, \alpha =1,2,3
\end{equation}
is conserved. Therefore the axial charge operator~:\begin{equation}
\label{Eq_LET.2}
Q_{5}(t)=\int d\vec{r}\, j^{5}_{0}(t,\vec{r})
\end{equation}
 is time independent. Due to the spontaneous breaking of the \( SU(2)_{L}\times SU(2)_{R} \)
symmetry down to \( SU(2)_{V} \), the axial charge does not anihilate
the vacuum state but instead creates states with zero momentum (massless)
pions. This is the origin of the soft pion theorems. The symmetry
is also explicitely broken by the small quark mass and for our purposes
it is sufficient to use the Partially Conseved Axial Current (PCAC)
formulation \cite{Adl68} which we write in the form~:\[
\partial ^{\mu }j_{\mu }^{5,\alpha }\approx m_{\pi }^{2}.\]
To derive the low soft pion theorems one does not need to be more
specific about the non conservation of the axial current\cite{Das67}.
One just needs to know the transformation properties of the relevant
operators under the symmetry group and that the symmetry breaking
is of order \( m_{\pi }^{2} \). To exploit these hypotheses one first
defines the iso-vector operator~:\cite{Furlan72}
\begin{equation}
\label{Eq_LET.8}
\bar{Q}=\left( Q_{5}+\frac{i}{m_{\pi }}\frac{d}{dt}Q_{5}\right) _{t=0},
\end{equation}
 which satisfies~:\begin{eqnarray}
<\pi ^{\beta }(K)|\bar{Q}^{\alpha }|0> & = & 0,\label{Eq_LET.9} \\
<0|\bar{Q}^{\alpha }|\pi ^{\beta }(K)> & = & 2if_{\pi }E_{\pi
}(K)(2\pi )^{3}\delta (\alpha ,\beta )\delta (\vec{K}),
\label{Eq_LET.10} 
\end{eqnarray}
where \( E_{\pi }(K)=\sqrt{\vec{K}^{2}+m_{\pi }^{2}} \) and \( f_{\pi } \)
is the pion decay constant. A general matrix element of \( \bar{Q} \)
obviously satisfies~:
\begin{eqnarray}
 &  & (E_{1}-E_{2})<p_{1}|\bar{Q}|p_{2}>=i\frac{E_{1}-E_{2}-m_{\pi }}{m_{\pi }}<p_{1}|\dot{Q}_{5}|p_{2}>\nonumber \\
 &  & =i(2\pi )^{3}\delta (\vec{p}_{1}-\vec{p}_{2})(E_{1}-E_{2}-m_{\pi
 })<p_{1}|\frac{\partial _{\mu }j^{\mu 5}(0)}{m_{\pi }}|p_{2}>.
\label{Eq_LET.11} 
\end{eqnarray}
 If one has \( E_{1}\neq E_{2} \) in the symmetry limit, then Eq.(\ref{Eq_LET.11})
amounts to~:\begin{equation}
\label{Eq_LET.11.1}
<p_{1}|\bar{Q}|p_{2}>=i(2\pi )^{3}\delta (\vec{p}_{1}-\vec{p}_{2})<p_{1}|\frac{\partial _{\mu }j^{\mu 5}(0)}{m_{\pi }}|p_{2}>,
\end{equation}
 which vanishes by the PCAC hypothesis. Therefore, if one evaluates
the matrix element of \( [\bar{Q},A] \) between nucleon states according
to~: \begin{equation}
\label{Eq_LET.12}
<N(p')|[\bar{Q},A]|N(p)>=\sum _{X}<N(p')|\bar{Q}|X><X|A|N(p)>-c.t.,
\end{equation}
 where the sum includes integration over the momenta and \( c.t. \)
denotes the crossed term, then the only states which survive in the
chiral limit are those for which~: \begin{equation}
\label{Eq_LET.13}
E_{X}=E_{N}(p)\, \, {\rm or}\, \, E_{X}=E_{N}(p').
\end{equation}
 This limits the sum over X to the following possibilities~:

\begin{enumerate}
\item X is the nucleon (Fig. \ref{Fig_soft_pion}a). 
\item X is the semi-disconnected \( \pi N \) state (Fig. \ref{Fig_soft_pion}b)with
the soft pion anihilated by \( \bar{Q} \) (the crossed term vanishes
due to Eq.(\ref{Eq_LET.9})). This is, up to a factor, the soft pion
production amplitude we are looking for.
\item X is the semi-disconnected \( \pi N \) state (Fig. \ref{Fig_soft_pion}c)
with the soft pion created or anihilated by \( A \).
\end{enumerate}
Other intermediate states, for instance an isobar replacing the nucleon
in Fig. \ref{Fig_soft_pion}a or a heavy meson replacing the pion
in Fig. \ref{Fig_soft_pion}c, cannot be degenerate with the
initial or final state and therefore are suppressed by PCAC. This
is also the case for pions loops some of which are shown on Fig. \ref{Fig_pion_loop}.
Here the energy gap is due to the fact that the intermediate pions
carry momentum. As is well known \cite{Lang73}, the vanishing chiral
limit of these contributions is in general reached in a non analytical
way due to the infra-red part of the momentum integration. The leading
non analytic term is due to the loop shown in Fig. \ref{Fig_pion_loop}a
and its infra-red part involves again the production of a soft pion
by the operator \( A \). So it could be computed within our framework
without extra hypothesis. However, at this stage of investigation
of DVCS, we think it is reasonnable to keep only the terms which do
not vanish in the chiral limit, i.e. those shown on Fig. \ref{Fig_soft_pion}.

We note that the term corresponding to Fig. \ref{Fig_soft_pion}c
does not contribute to pion electroproduction since the electromagnetic
current cannot create a pion out of the vacuum. By contrast it can
contribute to pion twist-2 production if the momentum of the intermediate
pion, that is \( \vec{q}-\vec{q}' \) in this case, is of order \( m_{\pi }. \)
Since \( t_{\gamma }=(q-q')^{2}<0 \), we have \( (\vec{q}-\vec{q}')^{2}>|t_{\gamma }| \)
and therefore we can neglect this term if we impose that \( |t_{\gamma }| \)
be much larger than \( m_{\pi }^{2} \). More precisely we ask that
\( t_{\gamma } \) or \( t \) be a typical hadronic scale which need
not be large but which does not vanishes when we let \( m_{\pi } \)
go to zero. Clearly this is compatible with the conditions (\ref{Eq_leading.0})
for factorisation of the DVCS reaction. Though this is not immediately
necessary \emph{we from now on assume that this condition is realised
independently of the reaction considered, that is whatever the operator
\( A. \) }

Under this condition we can write the soft pion expansion of the commutator
as~:\begin{eqnarray}
 &  & <N(p')|[\bar{Q}^{\alpha },A]|N(p)>=\nonumber \\
 &  & \int \frac{d\vec{K}}{(2\pi )^{3}2E_{N}(K)}\left( <N(p')|\bar{Q}^{\alpha }|N(K)><N(K)|A|N(p)>-c.t.\right) +\nonumber \\
 &  & \int \frac{d\vec{K}}{(2\pi )^{3}2E_{\pi }(K)}<0|\bar{Q}^{\alpha }|\pi ^{\beta }(K)><N(p')\pi ^{\beta }(K)|A|N(p)>+O(m_{\pi }).\, \, \, \label{Eq_LET.14} 
\end{eqnarray}
where the sum over the spin and isospin labels is understood. From
the definition of the isovector axial form factors~:\begin{equation}
\label{Eq_LET.15}
<N(p')|\bar{q}\gamma ^{\mu }\gamma _{5}\frac{\tau ^{\alpha }}{2}\Pi _{ud}q(0)|N(p)>=\bar{u}(p')\left[ g_{A}(t)\gamma ^{\mu }+h_{A}(t)\frac{(p'-p)^{\mu }}{2M}\right] \gamma _{5}\frac{\tau _{N}^{\alpha }}{2}u(p)
\end{equation}
we have~:\begin{equation}
\label{Eq_LET.16}
<N(p')|\bar{Q}^{\alpha }|N(p)>=(2\pi )^{3}\delta (\vec{p}\, '-\vec{p})g_{A}(0)u^{\dagger }(p)\gamma _{5}\frac{\tau ^{\alpha }_{N}}{2}u(p),
\end{equation}
 which, together with Eq.(\ref{Eq_LET.10}), leads to the following
expression of the commutator~:\begin{eqnarray}
<N(p')|[\bar{Q}^{\alpha },A]|N(p)> & = & g_{A}(0)\frac{1}{2E_{N}(p')}u^{\dagger }(p')\gamma _{5}\frac{\tau ^{\alpha }_{N}}{2}u(p')<N(p')|A|N(p)>\nonumber \\
 & - & g_{A}(0)\frac{1}{2E_{N}(p)}<N(p')|A|N(p)>u^{\dagger }(p)\gamma _{5}\frac{\tau ^{\alpha }_{N}}{2}u(p)\nonumber \\
 & + & if_{\pi }<N(p')\pi ^{\alpha }(m_{\pi },\vec{0})|A|N(p)>+O(m_{\pi }).\label{Eq_LET.17} 
\end{eqnarray}
 Let us now restore the spin label and define the on mass shell vertex
\( \Gamma  \) by~:%
\footnote{In the case of pion twist-2 production \( \Gamma  \) also depends
on \( x \) and on the Sudakov vector \( \tilde{n} \) but we omit
them as they are not important for what follows. 
} \begin{equation}
\label{Eq_LET.18}
<N(K',\sigma ')|A|N(K,\sigma )>=\bar{u}(K',\sigma ')\Gamma ([K'],[K])u(K,\sigma )
\end{equation}
where \( [K] \) is our special notation for the on shell 4-momentum
that is%
\footnote{By definition the argument of spinor is always on the mass shell so
we do not specify it.
}~: \begin{equation}
\label{Eq_LET.18.1}
[\vec{K}]=\vec{K},\, \, \, [K]_{0}=\sqrt{\vec{K}^{2}+M^{2}}.
\end{equation}
It is then a trivial task to show that Eq.(\ref{Eq_LET.17}) is, up
to terms of order \( m_{\pi }, \) equivalent to~:\begin{equation}
\label{Eq_LET.19}
<N(p')|[\bar{Q}^{\alpha },A]|N(p)>=-if_{\pi }T_{B}+if_{\pi }<N(p')\pi ^{\alpha }(m_{\pi },\vec{0})|A|N(p)>+O(m_{\pi }),
\end{equation}
with the Born term \( T_{B} \) defined as~: \begin{eqnarray}
T^{\alpha }_{B} & = & i\frac{g_{A}(0)}{2f_{\pi }}\bar{u}(p')\left[ \gamma .k_{\pi }\gamma _{5}\tau ^{\alpha }\frac{1}{\gamma .(p'+k_{\pi })-M}\Gamma ([p'+k_{\pi }],[p])\right. \nonumber \\
 & + & \left. \Gamma ([p'],[p-k_{\pi }])\frac{1}{\gamma .(p-k_{\pi })-M}\gamma .k_{\pi }\gamma _{5}\tau ^{\alpha }\right] u(p),\label{Eq_LET.20} 
\end{eqnarray}
and \( k_{\pi }=(m_{\pi },\vec{0}) \) is the 4-momentum of the pion
at rest. The expression (\ref{Eq_LET.20}) is formally covariant so
we can use it in a frame where the pion is not at rest. The \( O(m_{\pi }) \)
corrections to the soft pion expansion then become \( O(k_{\pi }) \)
where \( k_{\pi } \) is the 4-momentum of the moving pion. In
this way one generates correctly only the terms which go like the
velocity (\( \approx k_{\pi }/m_{\pi } \) ) of the pion. The other
terms of order \( k_{\pi } \) in (\ref{Eq_LET.20}) cannot be distinguished
from the other corrections to the soft pion expansion.

Note that the expression of the Born term \( T_{B} \) is not exactly
what one would expect from an effective Feynman diagram with pseudo-vector
\( \pi N \) coupling because, in this case, the vertex \( \Gamma  \)
would appear in the form \( \Gamma (p'+k_{\pi },p) \) or \( \Gamma (p',p-k_{\pi }) \).
This mismatch is due to the fact that in the soft pion expansion the
vertex \( \Gamma  \) comes naturally into play with its arguments
on the mass shell, and therefore cannot conserve energy. In practice
this energy non conservation is of order \( k_{\pi } \) so one could
replace \( [p'+k_{\pi }] \) and \( [p-k_{\pi }] \) by their off
mass shell values \( p'+k_{\pi } \) and \( p-k_{\pi } \) since this
would amount to change the \( O(k_{\pi }) \) correction. However
the advantage would be only of cosmetic nature because we do not know
what is \( \Gamma  \) when its arguments are off the mass shell!
So we shall retain expression (\ref{Eq_LET.20}).

\subsection{Evaluation of the commutators}
\label{subsec:comm}

In the symetry limit the axial charge is time independent and we now
that the symetry breaking term \( \dot{Q}_{5} \) vanishes as \( m_{\pi }^{2} \)
. So to compute the matrix element of the commutator~: \begin{equation}
\label{Eq_LET.21}
<N|[\bar{Q},A]|N>=<N|[Q_{5}+\frac{i}{m_{\pi }}\dot{Q}_{5},A]|N>,
\end{equation}
 one can use the commutations rules~: \begin{equation}
\label{Eq_LET.23}
[Q_{5}^{\alpha },q(x)]=-\gamma _{5}\frac{\tau ^{\alpha }}{2}\Pi _{ud}q(x),
\end{equation}
and neglect the term \( \dot{Q}_{5}/m_{\pi } \). The error will be
of order \( m_{\pi } \) unless there is a kinematical enhancement
due to the proximity of a pion pole in the matrix element (\ref{Eq_LET.21}).
To check this we look for such poles in the amplitudes and the dangerous
ones are shown on Fig. \ref{Fig_pi_pole}. In the case of electroproduction
the pion pole can only be in the \( N\bar{N} \) channel as shown
on Fig. \ref{Fig_pi_pole}a. In the case of associated DVCS it can
appear in the \( \gamma \gamma  \) channel (Fig. \ref{Fig_pi_pole}b),
in the \( N\bar{N} \) channel (Fig. \ref{Fig_pi_pole}c) or in the
\( (\gamma \pi ) \) channel (Fig. \ref{Fig_pi_pole}d). The \( \gamma \pi  \)
channel is suppressed by powers of \( 1/Q \) , so we are left with
pion poles which are of the form~:\[
\frac{1}{t-m_{\pi }^{2}}\, {\rm or}\, \frac{1}{t_{\gamma }-m_{\pi }^{2}},\]
 which are not dangerous since we have assumed that \( t \) or \( t_{\gamma } \)
remain finite in the chiral limit.%
\footnote{Note that the same argument has already been used to neglect the terms
shown on \ref{Fig_soft_pion}c in the saturation of the commutator.
Those terms would effectively contribute to the pion poles of Fig.
\ref{Fig_pi_pole}a and Fig. \ref{Fig_pi_pole}b.
} 

{
\begin{figure}
{\centering \resizebox*{0.6\textwidth}{!}{\includegraphics{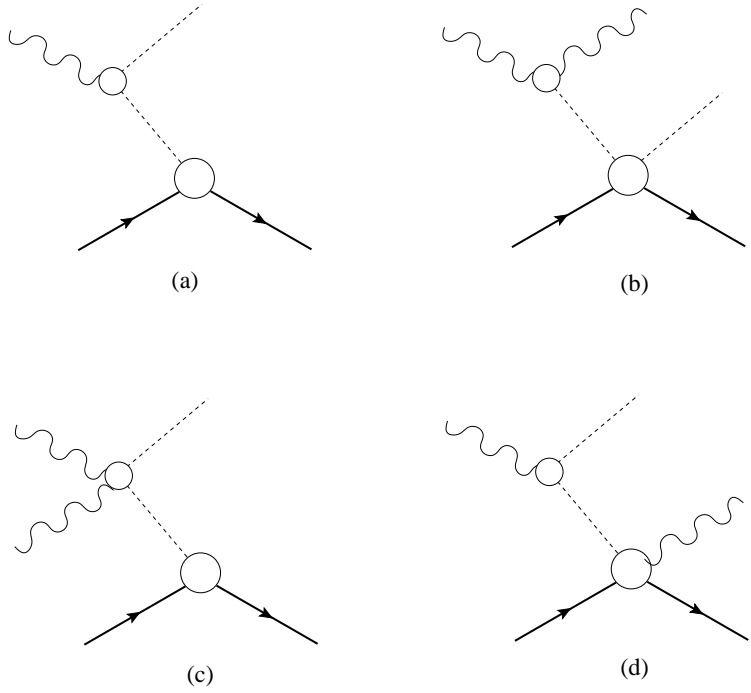}} \par}

\caption{\label{Fig_pi_pole}Pion poles in the amplitudes for electro-production
(a) and associated DVCS (b, c, d).}
\end{figure}
\par}
\vspace{0.3cm}

Using Eq.(\ref{Eq_LET.23}) it is straightforward to compute the relevant
commutators. From the definitions (\ref{Eq_prelim.3}, \ref{Eq_leading.8.2})
one gets~:\begin{eqnarray}
[Q_{5}^{\alpha },j^{\mu }] & = & i\varepsilon _{\alpha 3\beta }j^{\mu \beta }_{5},\label{Eq_LET.24} \\
\, [Q_{5}^{\alpha },\sum _{f}{\cal Q}_{f}^{2}S_{f}] & = & \frac{i}{3}\varepsilon _{\alpha 3\beta }S^{5}_{V\alpha },\label{Eq_LET.25} \\
\, [Q_{5}^{\alpha },\sum _{f}{\cal Q}_{f}^{2}S^{5}_{f}] & = & \frac{i}{3}\varepsilon _{\alpha 3\beta }S_{V\alpha },\label{Eq_LET.26} 
\end{eqnarray}
where \( S^{(5)}_{V\alpha } \) are the iso-vector twist-2 operators
introduced in (\ref{Eq_leading.8.1},\ref{Eq_leading.8.2}).

\subsection{Soft pion theorem for the pion electroproduction}

From Eqs.~(\ref{Eq_LET.19},\ref{Eq_LET.24}) we get the soft pion theorem,
which we write directly for a moving pion~: 
\begin{equation}
\label{Eq_LET.27}
<N(p')\pi ^{\alpha }(k_{\pi })|j^{\mu }|N(p)>=T_{B}+\frac{1}{f_{\pi }}\varepsilon _{\alpha 3\beta }<N(p')|j^{\mu \beta }_{5}|N(p)>+O(k_{\pi }),
\end{equation}
where the Born term is obtained from Eq.(\ref{Eq_LET.20}) with~: \begin{eqnarray}
\Gamma ([K'],[K]) & = & \left( F_{1}^{p}([t])\frac{1+\tau _{3}^{N}}{2}+F_{1}^{n}([t])\frac{1-\tau _{3}^{N}}{2}\right) \gamma ^{\mu }\nonumber \\
 & + & i\left( F_{2}^{p}([t])\frac{1+\tau _{3}^{N}}{2}+F_{2}^{n}([t])\frac{1-\tau _{3}^{N}}{2}\right) \frac{\sigma ^{\mu \nu }([K']-[K])_{\nu
 }}{2M},\, \, \label{Eq_LET.28} 
\end{eqnarray}
and \( [t]=([K']-[K])^{2}. \) This is nothing but the low energy
theorem originally derived by Nambu \emph{et al.} \cite{Nambu62a,Nambu62b}.
Using our assumption that \( t \) is non zero in the soft pion limit,
one can check that the amplitude (\ref{Eq_LET.27}) respects gauge
invariance in the form \begin{equation}
\label{Eq_LET.29}
(p'+k_{\pi }-p)_{\mu }\left( T_{B}+\frac{1}{f_{\pi }}\varepsilon _{\alpha 3\beta }<N(p')|j^{\mu \beta }_{5}|N(p)>+O(k_{\pi })\right) =0,
\end{equation}
 that is the additional terms needed to have exact gauge invariance
are in the \( O(k_{\pi }) \) corrections
\footnote{That would not be true for instance in pion photo-production because
at threshold \( t \) is of order \( m_{\pi }^{2} \)
}. The axial current matrix element which appears in Eq.(\ref{Eq_LET.27})
is given by Eq.(\ref{Eq_LET.15}) and involves the form factors \( g_{A}(t) \)
and \( h_{A}(t) \). The latter is not so well known experimentally
but this does not matter because it is multiplied by \( p'-p=k_{\gamma }-k_{\pi } \)
where \( k_{\gamma }=q-q' \) is the photon exchanged in the BH process.
In the Lorentz gauge the term proportional to \( k_{\gamma } \) does
not contribute to the amplitude and, because \( t \) does not vanish
in the soft pion limit, the term linear in \( k_{\pi } \) can be
pushed in the \( O(k_{\pi }) \) corrections despite the pion pole
in \( h_{A}(t) \). The rest of the calculation is just algebraic
manipulation of expressions (\ref{Eq_LET.27},\ref{Eq_LET.28}) so
we do not need to give the details. For completeness we just mention
the relation between the charged and cartesian pion production amplitudes~:\begin{equation}
\label{Eq_LET.30}
T(\pi ^{\pm })=\frac{T(\pi ^{1}\mp i\pi ^{2})}{\sqrt{2}},\, \, \, T(\pi ^{0})=T(\pi ^{3}).
\end{equation}

\subsection{Soft pion theorem for the pion twist-2 production\label{LET_twist2}}

The soft pion theorems for pion twist-2 production folllow from Eqs(\ref{Eq_LET.19},\ref{Eq_LET.25},\ref{Eq_LET.26})~:\begin{eqnarray}
 &  & <N(p')\pi ^{\alpha }(k_{\pi })|\sum _{f}{\cal Q}_{f}^{2}S_{f}(x,\tilde{n})|N(p)>=\nonumber \\
 &  & T_{B}+\frac{1}{3f_{\pi }}\varepsilon _{\alpha 3\beta }<N(p')|S^{5}_{V\beta }(x,\tilde{n})|N(p)>+O(k_{\pi }),\label{Eq_LET.31} \\
 &  & <N(p')\pi ^{\alpha }(k_{\pi })|\sum _{f}{\cal Q}_{f}^{2}S^{5}_{f}(x,\tilde{n})|N(p)>=\nonumber \\
 &  & T^{5}_{B}+\frac{1}{3f_{\pi }}\varepsilon _{\alpha 3\beta }<N(p')|S_{V\beta }(x,\tilde{n})|N(p)>+O(k_{\pi }),\label{Eq_LET.32} 
\end{eqnarray}
where the Born terms \( T^{(5)}_{B} \) are obtained from Eq.(\ref{Eq_LET.20})
with the vertices \( \Gamma ^{(5)} \) defined by (see Eq.(\ref{Eq_LET.18}))~:
\begin{eqnarray}
<N(K')|\sum _{f}{\cal Q}_{f}^{2}S_{f}(x,\tilde{n})|N(K)> & = & \bar{u}(K')\Gamma ([K'],[K],x,\tilde{n})u(K),\label{Eq_LET.33} \\
<N(K')|\sum _{f}{\cal Q}_{f}^{2}S^{5}_{f}(x,\tilde{n})|N(K)> & = & \bar{u}(K')\Gamma ^{5}([K'],[K],x,\tilde{n})u(K).\label{Eq.LET.34} 
\end{eqnarray}
Note that we have restored the explicit dependance on \( x,\tilde{n} \).
Using the parametrisation (\ref{Eq_leading.9.2}) we have~:\begin{eqnarray}
 &  & \Gamma ([K'],[K],x,\tilde{n})=\frac{2}{([K]+[K']).\tilde{n}}\times \nonumber \\
 &  & \sum _{f}{\cal Q}_{f}^{2}\left( H_{f/N}([x],[\xi ],[\Delta ]^{2})\gamma .\tilde{n}+iE_{f/N}([x],[\xi ],[\Delta ]^{2})\frac{\sigma ^{\mu
 \nu }\tilde{n}_{\mu }[\Delta ]_{\nu }}{2M}\right), \label{Eq_LET.35} \\
 &  & \Gamma ^{5}([K'],[K],x,\tilde{n})=\frac{2}{([K]+[K']).\tilde{n}}\times \nonumber \\
 &  & \sum _{f}{\cal Q}_{f}^{2}\left( \tilde{H}_{f/N}([x],[\xi ],[\Delta ]^{2})\gamma .\tilde{n}+\tilde{E}_{f/N}([x],[\xi ],[\Delta
 ]^{2})\frac{\tilde{n}_{.}[\Delta ]}{2M}\right) \gamma ^{5},\label{Eq_LET.36} 
\end{eqnarray}
where the bracketed variables are, by definition (see Section \ref{GPD})~:\[
[x]=2\frac{x}{([K]+[K']).\tilde{n}},\, \, \, [\xi ]=-\frac{[\Delta ].\tilde{n}}{([K]+[K']).\tilde{n}},\, \, \, [\Delta ]=[K']-[K].\]
From Eq.(\ref{Eq_LET.20}) we see that one has either \( ([K]=[p],\, [K']=[p'+k_{\pi }]) \)
or \( ([K]=[p-k_{\pi }],\, [K']=[p']) \). It is then apparent that
the approximation \( ([x]\rightarrow x\, [\xi ]\rightarrow \xi ,\, [\Delta ]\rightarrow \Delta ) \)
in Eqs(\ref{Eq_LET.35},\ref{Eq_LET.36}) only induces corrections
of order \( k_{\pi } \). Similarly the matrix elements of \( S_{V}^{(5)} \)
which appear in Eqs.(\ref{Eq_LET.31},\ref{Eq_LET.32}) can be written
as~: \begin{eqnarray}
 &  & <N(p')|S_{V\alpha }(x,\tilde{n})|N(p)>=\frac{2}{(p+p').\tilde{n}}\times \nonumber \\
 &  & \bar{u}(p')\left[ H_{V}(\hat{x},\hat{\xi },t)\gamma .\tilde{n}+iE_{V}(\hat{x},\hat{\xi },t)\frac{\sigma ^{\mu \nu }\tilde{n}_{\mu }(p'-p)_{\nu }}{2M}\right] \frac{\tau _{\alpha }^{N}}{2}u(p),\label{Eq_LET.37} \\
 &  & <N(p')|S^{5}_{V\alpha }(\kappa ,\tilde{n})|N(p)>=\frac{2}{(p+p').\tilde{n}}\times \nonumber \\
 &  & \bar{u}(p')\left[ \tilde{H}_{V}(\hat{x},\hat{\xi },t)\gamma .\tilde{n}+\tilde{E}_{V}(\hat{x},\hat{\xi
 },t)\frac{\tilde{n}.(p'-p)}{2M}\right] \frac{\tau _{\alpha }^{N}}{2}u(p),\label{Eq_LET.38} 
\end{eqnarray}
where now the hatted variables are~:\[
\hat{x}=2\frac{x}{(p+p').\tilde{n}},\, \, \, \hat{\xi }=-\frac{(p'-p).\tilde{n}}{(p+p').\tilde{n}},\]
and \( t=(p'-p)^{2}. \) Again the approximation \( \hat{x}\rightarrow x,\, \hat{\xi }\rightarrow \xi ,\, (p'-p)\rightarrow \Delta  \)
induces only corrections of order \( k_{\pi } \). This completes
our derivation of the soft pion theorem for twist-2 production.

A few comments are in order. First we see that, as promised in the
introduction, the soft pion theorem for twist-2 production involves
no new quantities with respect to the elastic case. Second we note
that the commutator term, which involves only the isovector operators
\( S_{V}^{(5)} \), depends essentially on the valence quarks while
the Born term, which sums over all favours, is sensitive both to the
valence and the sea quarks. Third we recall that gauge invariance
is not an issue for the pion twist-2 production. The couplings to
the photon fields have been factorized from the beginning, see Eq.(\ref{Eq_leading.6}),
independently of the final hadronic state. As in the elastic case,
gauge invariance can be restored by a term which is explicitely of
higher order in \( t/Q \) , and thus does not contribute in the Bjorken
limit \cite{Vdh99}.

\section{Estimate of the \protect\( \Delta (1232)\protect \) contribution\label{ADVCS_Delta}}

The soft pion theorem proposed in the previous section can be valid
only for small values of the pion momentum. One can expect strong
deviations when the mass of the pion-nucleon system approaches the
first nucleon excitation, that is the \( \Delta (1232). \) So to
improve the range of validity of our estimates we now calculate the
contribution of this resonance. 
\begin{figure}
{\centering \resizebox*{0.6\textwidth}{!}{\includegraphics{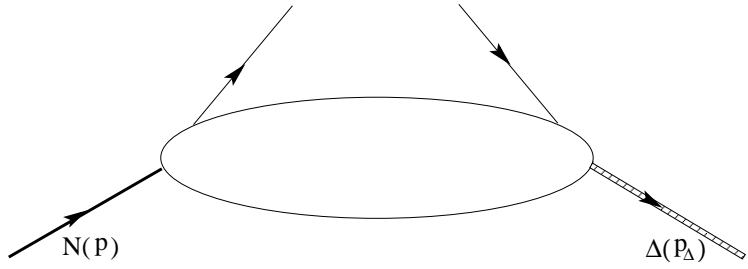}} \par}

\caption{\label{Fig_twist2_delta}Delta excitation by the twist-2 operator.}
\end{figure}
As a starting point, we consider the matrix element as shown in Fig.~\ref{Fig_twist2_delta},
where the \( \Delta  \)-resonance is on its mass-shell. Subsequently,
we discuss the modification due to the \( \Delta \rightarrow \pi N \)
strong decay.

The \( N\rightarrow \Delta  \) matrix element for the vector twist-2
operator has the following parametrization~\cite{Fra00,GPV01} 
\footnote{Note that there are four $N \to \Delta$ helicity
amplitudes for the vector operator. However,   
gauge invariance for the electromagnetic current operator leads to
only three electromagnetic form factors. Hence there are
strictly speaking four GPDs for the vector $N \to \Delta$ transition, 
of which one has a vanishing first moment. 
In view of our strategy to keep only the dominant transitions, we will
neglect this GPD which has a vanishing first moment in the following.}~:
\begin{eqnarray}
&&<\Delta (p_{\Delta })|\sum _{f}{\cal Q}_{f}^{2}S_{f}(x,\,\tilde{n})|N(p)> 
\nonumber \\
& = & {1\over 6}\; \bar{\psi }^{\beta }(p_{\Delta })\; T^{\dagger}_{3} \; 
\left\{ \, H_{M}(x,\xi ,\Delta^2)\; 
\left( -{\mathcal{K}}_{\beta \kappa }^{M} \right) \, \tilde n^{\kappa }\right. 
\;+\; H_{E}(x,\xi ,\Delta^2)\; 
\left( -{\mathcal{K}}_{\beta \kappa }^{E} \right) \, \tilde n^{\kappa } \nonumber \\
&&\hspace{2.5cm} \left. +\; H_{C}(x,\xi ,\Delta^2)\; 
\left(- {\mathcal{K}}_{\beta \kappa }^{C} \right) \, \tilde n^{\kappa }\right\} 
\, u(p), 
\label{eq:ndelvec} 
\end{eqnarray}
where \( \psi ^{\beta }(p_{\Delta }) \) is the Rarita-Schwinger spinor
for the \( \Delta  \)-field and \( T^{\dagger }_{3} \) is the isospin
\( 1/2\rightarrow 3/2 \) transition operator, satisfying~: \begin{eqnarray}
\langle \frac{3}{2}T_{z}\; |\; T^{\dagger }_{\lambda }\; |\; \frac{1}{2}t_{z}\rangle \; =\; \langle \frac{1}{2}t_{z}\; ,\; 1\lambda \; |\; \frac{3}{2}T_{z}\rangle \, ,\label{Eq_DEL.2} 
\end{eqnarray}
 with \( t_{z} \) (\( T_{z} \)) the isospin projections of \( N \)
(\( \Delta  \)) respectively, and where \( \lambda =0,\pm 1 \) indicate
the spherical components of the isovector operator \( T^{\dagger } \).
The factor 1/6 in Eq.~(\ref{eq:ndelvec}) results from the quadratic
quark charge combination \( (e_{u}^{2}-e_{d}^{2})/2 \). Furthermore,
in Eq.~(\ref{eq:ndelvec}), the covariants \( {\mathcal{K}}^{M,E,C}_{\beta \mu } \)
are the magnetic dipole, electric quadrupole, and Coulomb quadrupole
covariants \cite{Jon73}~:

\begin{eqnarray}
{\mathcal{K}}_{\beta \mu }^{M} & = & -i\frac{3(M_{\Delta }+M)}{2M((M_{\Delta }+M)^{2}-\Delta^2)}\varepsilon _{\beta \mu \lambda \sigma }P^{\lambda }\Delta ^{\sigma }\; ,\nonumber \\
{\mathcal{K}}_{\beta \mu }^{E} & = & -{\mathcal{K}}_{\beta \mu }^{M}-\frac{6(M_{\Delta }+M)}{MZ(\Delta^2)}\varepsilon _{\beta \sigma \lambda \rho }P^{\lambda }\Delta ^{\rho }\varepsilon ^{\sigma }_{\, \, \mu \kappa \delta }P^{\kappa }\Delta ^{\delta }\gamma ^{5}\; ,\label{K-def} \\
{\mathcal{K}}_{\beta \mu }^{C} & = & 
-i\frac{3(M_{\Delta}+M)}{MZ(\Delta^2)}\Delta _{\beta }(\Delta^2 P_{\mu }-\Delta \cdot P\Delta _{\mu })\gamma ^{5}\; ,\nonumber 
\end{eqnarray}
where we introduced the notation~:\begin{equation}
\label{Eq_DEL.4}
Z(\Delta^2)=[(M_{\Delta }+M)^{2}-\Delta^2][(M_{\Delta }-M)^{2}-\Delta^2].
\end{equation}
Here \( P=(p_{\Delta }+p)/2 \), \( \Delta =p_{\Delta }-p \), 
and \( p_{\Delta }^{2}=M_{\Delta }^{2} \) (\( M_{\Delta } \) = 1232
MeV). 
As we are considering the production of a $\Delta$ here, we
explicitely adopt the notation $\Delta^2$ in all expressions in this
section to denote the momentum transfer squared to the hadronic system, 
in order not to confuse with the variable $t$ introduced before.  
In Eq.~(\ref{eq:ndelvec}), the GPDs \( H_{M} \), \( H_{E} \),
and \( H_{C} \) for the \( N\rightarrow \Delta  \) vector transition
are linked with the three \( N\rightarrow \Delta  \) vector current
transition form factors \( G_{M}^{*} \), \( G_{E}^{*} \), and \( G_{C}^{*} \)
through the sum rules~: \begin{eqnarray}
\int _{-1}^{1}dx\, \, H_{M,E,C}(x,\xi ,\Delta^2)
=2\, \, G_{M,E,C}^{*}(\Delta^2)\; ,
\label{Eq_DEL.5} 
\end{eqnarray}
 where \( G_{M,E,C}^{*}(\Delta^2) \) are the standard magnetic dipole, electric
quadrupole and Couloub quadrupole transition form factors respectively
\cite{Jon73}. As is well known from experiment, the \( N\rightarrow \Delta  \)
vector transition at small and intermediate momentum transfers is
largely dominated by the \( N\rightarrow \Delta  \) 
magnetic dipole excitation, parametrized by \( G_{M}^{*}(\Delta^2)\). 
At \( \Delta^2=0 \), its value extracted
from pion photoproduction experiments is given by \( G_{M}^{*}(0)\simeq 3.02 \)
\cite{Tia00}. For its \( \Delta^2 \)-dependence, 
we use a recent phenomenological
parametrization \cite{Tia00} from a fit to pion electroproduction
data. Given the smallness of the electric and Coulomb quadrupole 
\( N\rightarrow \Delta  \)
transitions, we will neglect in the following the contribution of
the GPDs \( H_{E} \) and \( H_{C} \). 

The \( N\rightarrow \Delta  \) matrix element for the axial twist-2
operator has the following parametrization~\cite{Fra00,GPV01}~:

\begin{eqnarray}
&&<\Delta (p_{\Delta })|\sum _{f}{\cal Q}_{f}^{2}S_{f}^{5}(x,\,
\tilde{n})|N(p)> \nonumber \\
&=& 
{1\over 6}\; \bar{\psi }^{\beta }(p_{\Delta })\; T^{\dagger }_{3} \;
\sqrt{3 \over 2} \,
\left\{ C_{1}(x,\xi ,\Delta^2)\, \tilde n_{\beta }\, 
+\, C_{2}(x,\xi ,\Delta^2)\, \Delta _{\beta }\, {{\Delta \cdot \tilde n}\over {M^{2}}}
\right. \nonumber \\
 &  & \hspace{3cm} +\, C_{3}(x,\xi ,\Delta^2)\, {1\over {M}}\left(
{\gamma .\Delta \, \tilde n_{\beta }-{\gamma . \tilde n}\, \Delta _{\beta }}\right) \nonumber \\
 &  & \hspace{3cm} \left. +\, C_{4}(x,\xi ,\Delta^2)\, {2\over {M^{2}}}
\left( {\Delta \cdot P}\, \tilde n_{\beta }-\Delta _{\beta }\right) \right\} u(p) \; .
\label{eq:ndelax} 
\end{eqnarray}
The GPDs \( C_{1},C_{2},C_{3} \) and \( C_{4} \) entering in 
Eq.~(\ref{eq:ndelax})
are linked with the four \( N\rightarrow \Delta  \) axial-vector
current transition form factors 
\( C_{5}^{A}(\Delta^2) \), \( C_{6}^{A}(\Delta^2) \),
\( C_{3}^{A}(\Delta^2) \) and \( C_{4}^{A}(\Delta^2) \), 
introduced by Adler~\cite{Adl75}
through the sum rules 
\footnote{Note that the factor $\sqrt{3 \over 2}$ is conventionally
chosen in Eq.~(\ref{eq:ndelax})
such that the Adler form factors $C_i^A(\Delta^2)$ correspond
with the \(p \rightarrow \Delta ^{+} \) transition.}
~: 
\begin{eqnarray}
\int _{-1}^{1}dx\, C_{1}(x,\xi ,\Delta^2)\,  
& = & \, 2\, \, C_{5}^{A}(\Delta^2)\, ,\nonumber \\
\int _{-1}^{1}dx\, C_{2}(x,\xi ,\Delta^2)\,  
& = & \, 2\, \, C_{6}^{A}(\Delta^2)\, ,\nonumber \\
\int _{-1}^{1}dx\, C_{3}(x,\xi ,\Delta^2)\,  
& = & \, 2\, \, C_{3}^{A}(\Delta^2)\, ,\nonumber \\
\int _{-1}^{1}dx\, C_{4}(x,\xi ,\Delta^2)\,  
& = & \, 2\, \, C_{4}^{A}(\Delta^2)\, .
\label{eq:axial-sumrules} 
\end{eqnarray}
 For small momentum transfers \( \Delta^2 \), PCAC leads to a dominance
of the form factors \( C_{5}^{A} \) and \( C_{6}^{A} \). For \( C_{5}^{A} \),
a Goldberger-Treiman relation for the \( N\rightarrow \Delta  \)
transition leads to~: 
\begin{eqnarray}
\sqrt{3 \over 2} \, C_{5}^{A}(0)\, =\, 
\frac{f_{\pi N\Delta }}{2f_{\pi NN}}\, g_{A}\, .
\label{Eq_DEL.6} 
\end{eqnarray}
 Using the phenomenological values \( f_{\pi N\Delta }\simeq 1.95 \),
\( f_{\pi NN}\simeq 1.00 \), and \( g_{A}\simeq 1.267 \) one obtains \( C_{5}^{A}(0)\simeq 1.01 \). 
The form factor \( C_{6}^{A}(\Delta^2) \) at small values of 
\( \Delta^2 \)
is dominated by the pion-pole contribution, given by~: 
\begin{eqnarray}
C_{6}^{A}(\Delta^2)\, =\, 
\frac{M^{2}}{m_{\pi }^{2}-\Delta^2}\, C_{5}^{A}(0)\, .
\label{Eq_DEL.7} 
\end{eqnarray}
Because we are only interested in the limit $-t, -t_\gamma >> m_\pi^2$, 
we neglect the pion pole contribution of $C_6$ in the following,
consistent with the discussion of Section~\ref{subsec:comm}.
\newline
\indent
For the two remaining Adler form factors \( C_{3}^{A} \) and \( C_{4}^{A} \),
a detailed comparison with experimental data for \( \nu  \)-induced
\( \Delta ^{++} \) production led to the values \cite{Kit90}~:
\( C_{3}^{A}(0)\simeq 0.0 \) and \( C_{4}^{A}(0)\simeq -0.3 \).
Given the smallness of these values, we will neglect in the following
the contributions from the GPDs \( C_{3} \) and \( C_{4} \).

To provide numerical estimates for the \( N\rightarrow \Delta  \)
DVCS amplitudes, we need a model for the three remaining 'large' GPDs
which appear in Eqs.~(\ref{eq:ndelvec},\ref{eq:ndelax}) i.e. \( H_{M} \),
\( C_{1} \) and \( C_{2} \). Here we will be guided by the large
\( N_{c} \) relations discussed in~\cite{Fra00,GPV01}. These relations
connect the \( N\rightarrow \Delta  \) GPDs \( H_{M} \), \( C_{1} \),
and \( C_{2} \) to the \( N\rightarrow N \) isovector GPDs \( E^{u}-E^{d} \),
\( \tilde{H}^{u}-\tilde{H}^{d} \), and \( \tilde{E}^{u}-\tilde{E}^{d} \)
respectively. These relations are given by %
\footnote{Note that all other (sub-dominant) GPDs vanish at leading order in
the 1/\( N_{c} \) expansion.
}~:

\begin{eqnarray}
H_{M}(x,\xi ,\Delta^2) & = & 
\frac{2}{\sqrt{3}}\left[ E^{u}(x,\xi ,\Delta^2)
-E^{d}(x,\xi ,\Delta^2)\right] \; ,
\nonumber \\
C_{1}(x,\xi ,\Delta^2) & = & 
\sqrt{3}\left[ \tilde{H}^{u}(x,\xi ,\Delta^2)
-\tilde{H}^{d}(x,\xi ,\Delta^2)\right] \, ,\nonumber \\
C_{2}(x,\xi ,\Delta^2) & = & 
\frac{\sqrt{3}}{4}\left[ \tilde{E}^{u}(x,\xi ,\Delta^2)
-\tilde{E}^{d}(x,\xi ,\Delta^2)\right] \, .
\label{Eq_DEL.8} 
\end{eqnarray}

To give an idea of the accuracy of these relations, we calculate the
first moment of both sides of Eq.~(\ref{Eq_DEL.8}) and compares
their values at \( \Delta^2=0 \). 
For \( H_{M} \), we obtain for the \textit{lhs}
the phenomenological value \( 2G_{M}^{*}(0)\simeq 6.04 \), in comparison
with the large \( N_{c} \) prediction of \( (2/\sqrt{3})\kappa ^{V}\simeq 4.27 \),
accurate at the 30 \% level 
\footnote{Note that in the large $N_c$ limit, the isovector
combination $H^u - H^d$ is suppressed, therefore one could as well
give as estimate 
$2 G_{M}^{*}(0)\simeq {2 \over {\sqrt{3}}} \, \mu^V \simeq 5.43 $, 
which is accurate at the 10 \% level.}. 

For \( C_{1} \), the phenomenological
value yields \( 2C_{5}(0)\simeq 2.02 \), whereas the large \( N_{c} \)
prediction yields \( \sqrt{3}g_{A}\simeq 2.19 \), accurate at the
10\% level. For the pion-pole contribution to \( C_{6} \), we obtain
the same accuracy as for \( C_{5} \). 
Furthermore, for the \( N\rightarrow \Delta  \)
DVCS in the near forward direction, unlike the \( N\rightarrow N \)
DVCS case, the axial transition (proportional to the GPDs \( C_{1} \) )
is numerically more important than the vector transition (parametrized
by \( H_{M} \)). This is because \( H_{M} \) is accompanied by a
momentum transfer \( \Delta  \) in the tensor \( {\mathcal{K}}^{M} \)
as is seen from Eq.~(\ref{K-def}), in contrast with the structure
associated with the GPD \( C_{1} \). From these considerations, we
estimate that the large \( N_{c} \) relations of Eq.~(\ref{Eq_DEL.8})
for the \( N\rightarrow \Delta  \) transition allow us to estimate
at the \( \gamma ^{*}p\rightarrow \gamma \Delta  \) DVCS process,
at the 30 \% accuracy level or better.

In order to calculate the coherent sum of the non-resonant ADVCS process
discussed before and the \( \Delta  \)-resonant process, we next
discuss the modification of the \( \gamma ^{*}N\rightarrow \gamma \Delta  \)
matrix elements due to the \( \Delta \rightarrow \pi N \) strong
decay.

The matrix element of the vector operator for the \( \gamma ^{*}N\rightarrow \gamma \Delta  \)
transition followed by \( \Delta \rightarrow \pi N \) decay, is modified
from Eq.~(\ref{eq:ndelvec}) to~:
\begin{eqnarray}
 & & <\pi N|\sum _{f}{\cal Q}_{f}^{2}S_{f}(x,\, \tilde{n})|N>_{Delta} 
\nonumber \\
 & = & - \; {{\mathcal{I}}}\frac{f_{\pi N\Delta }}{m_{\pi }}\, 
{k_{\pi }}^{\alpha }\, \bar{u}(p')\frac{i\left( \gamma .p_{\Delta }+W\right) }{W^{2}-M_{\Delta }^{2}+iW\Gamma _{\Delta }(W)}\nonumber \\
 & \times  & \left[ g_{\alpha \beta }-\frac{1}{3}\gamma _{\alpha }\gamma _{\beta }-\frac{1}{3W}\left( \gamma _{\alpha }(p_{\Delta })_{\beta }-\gamma _{\beta }(p_{\Delta })_{\alpha }\right) -\frac{2}{3W^{2}}(p_{\Delta })_{\alpha }(p_{\Delta })_{\beta }\right] \nonumber \\
 & \times  & \, {1\over 6}\, \left\{ \, H_{M}(x,\xi ,\Delta^2)\; 
\left(- {\mathcal{K}}^{M}\right)^{\beta \kappa }\, \tilde n_{\kappa }\; 
+\; ...\right\} \, u(p),
\label{eq:ndelvec2} 
\end{eqnarray}
where \( p_{\Delta }\equiv p'+k_{\pi } \), \( \Delta \equiv p'+k_{\pi }-p \),
and where we only indicated the leading transition, proportional to
\( H_{M} \). The isospin factor \( {\mathcal{I}} \) takes on the
values : \( {\mathcal{I}}=2/3 \) for the \( \pi ^{0}p \) final state,
and \( {\mathcal{I}}=-\sqrt{2}/3 \) for the \( \pi ^{+}n \) final
state. In the \( \Delta  \) propagator in Eq.~(\ref{eq:ndelvec2}),
the energy-dependent width \( \Gamma _{\Delta }(W) \) is given by~:\begin{equation}
\label{Eq_DEL.9}
\Gamma _{\Delta }(W)\, =\, \Gamma _{\Delta }(M_\Delta)\; \cdot \;
\frac{\left| \vec{k}^{*}_{\pi }(W) \right| ^{3}}
{\left|\vec{k}^{*}_{\pi }(M_\Delta) \right| ^{3}} \;\cdot \;
\frac{M_\Delta}{W} \, ,
\end{equation}
with $\Gamma _{\Delta }(M_\Delta) \simeq 0.120$~GeV,  
and where \( |\vec{k}_{\pi }^{*}(W)| \) is given in Eq.~(\ref{Eq_prelim.9}).
\newline
\indent
Analogously, the matrix element of the axial-vector operator for the
\( \gamma ^{*}N\rightarrow \gamma \Delta  \) transition followed
by \( \Delta \rightarrow \pi N \) decay, is modified from Eq.~(\ref{eq:ndelax})
to~:\begin{eqnarray}
 & & <\pi N|\sum _{f}{\cal Q}_{f}^{2}S_{f}^{5}(x,\,\tilde{n})|N>_{Delta}
\nonumber \\
 & = & - \; {{\mathcal{I}}}\frac{f_{\pi N\Delta }}{m_{\pi }}
\, {k_{\pi }}^{\alpha }\, \bar{u}(p')\frac{i\left( \gamma .p_{\Delta }+W\right) }{W^{2}-M_{\Delta }^{2}+iW\Gamma _{\Delta }(W)}\nonumber \\
 & \times  & \left[ g_{\alpha \beta }-\frac{1}{3}\gamma _{\alpha }\gamma _{\beta }-\frac{1}{3W}\left( \gamma _{\alpha }(p_{\Delta })_{\beta }-\gamma _{\beta }(p_{\Delta })_{\alpha }\right) -\frac{2}{3W^{2}}(p_{\Delta })_{\alpha }(p_{\Delta })_{\beta }\right] \nonumber \\
 & \times  & \, {1\over 6}\, \sqrt{3 \over 2} \; 
\left\{ \, C_{1}(x,\xi ,\Delta^2)\, \tilde n^{\beta }
\; +\; ...\right\} \, u(p) ,
\label{eq:ndelaxvec2} 
\end{eqnarray}
where we again only indicated the leading transition proportional
to the GPD \( C_{1} \).

Finally, in order to calculate the \( \gamma ^{*}p\rightarrow \gamma \pi N \)
process, we also need to specify the BH process of Eq.~(\ref{Eq_prelim.1.2})
associated with the \( N\rightarrow \Delta  \) transition. The corresponding
\( \Delta  \) contribution to the matrix element \( J_{\nu } \)
entering in Eq.~(\ref{Eq_prelim.1.2}) is given by~: \begin{eqnarray}
 &   & <\pi N|j^{\nu }(0)|N>_{Delta}\nonumber \\
&\hspace{-1.25cm}  = & \hspace{-0.75cm} - \; {{\mathcal{I}}}\frac{f_{\pi N\Delta }}{m_{\pi }}\, {k_{\pi }}^{\alpha }\, \bar{u}(p')\frac{i\left( \gamma \cdot p_{\Delta }+W\right) }{W^{2}-M_{\Delta }^{2}+iW\Gamma _{\Delta }(W)}\nonumber \\
 & \hspace{-1.25cm} \times  & \hspace{-0.75cm} \left[ g_{\alpha \beta }-\frac{1}{3}\gamma _{\alpha }\gamma _{\beta }-\frac{1}{3W}\left( \gamma _{\alpha }(p_{\Delta })_{\beta }-\gamma _{\beta }(p_{\Delta })_{\alpha }\right) -\frac{2}{3W^{2}}(p_{\Delta })_{\alpha }(p_{\Delta })_{\beta }\right] \nonumber \\
&\hspace{-1.25cm} \times &  \hspace{-0.75cm} \left\{ \, G_{M}^{*}(\Delta^2)
\left(-{\mathcal{K}}^{M}\right) ^{\beta \nu }
+ G_{E}^{*}(\Delta^2)
\left( -{\mathcal{K}}^{E}\right) ^{\beta \nu }
 + G_{C}^{*}(\Delta^2) 
\left( -{\mathcal{K}}^{C}\right) ^{\beta \nu }\right\} \, u(p),
\label{eq:bhndelta} 
\end{eqnarray}
 in terms of the same vector \( N\rightarrow \Delta  \) transition
form factors \( G_{M,E,C}^{*} \) as discussed before. The isospin
factor \( {\mathcal{I}} \) is the same as specified following 
Eq.~(\ref{eq:ndelvec2}).
\section{Results\label{Results}}

\begin{figure}[ht]
{\centering
\resizebox*{0.85\textwidth}{!}{\includegraphics{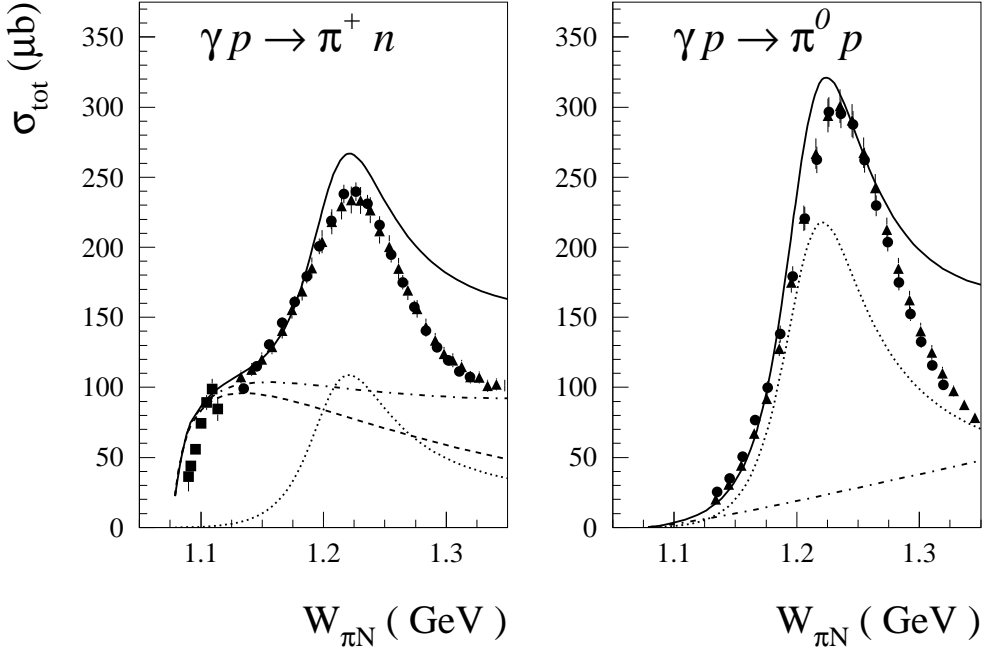}} \par}
\vspace{-0.5cm}
\caption{Total pion photoproduction cross sections for the different
model contributions considered in this paper. 
Dashed curve : commutator contribution. 
Dashed-dotted curves : commutator + Born contributions. 
Dotted curves : $\Delta$ contribution. 
Solid curves : commutator + Born + $\Delta$ contributions. 
The data are from Ref.~\cite{pitot_mcph} (diamonds), 
Ref.~\cite{pitot_macc} (circles), 
and Ref.~\cite{pitot_ahr} (triangles).
}
\label{fig:tot_gapi}
\end{figure}

In the absence of available data for the $\gamma^* p \to \gamma \pi N$
process, we can get an idea on the accuracy of our estimates  
by comparing the pion production amplitudes of
Eqs.~(\ref{Eq_LET.27},\ref{eq:bhndelta}), which enter in the
Bethe-Heitler process, with the pion photo- and electroproduction cross
sections. We are interested here in the region of not too large
$-t_\gamma < 1$ GeV$^2$, 
corresponding with the virtuality of the photon in the Bethe-Heitler
process. In this kinematical range, a large amount of pion photo- and
electroproduction data exist to compare with.  
In particular, the pion photoproduction cross section $\gamma p \to \pi N$  (proportional 
to the cross section of the associated Bethe-Heitler (ABH) 
process for $t_\gamma \to 0$), is given by
\begin{eqnarray}
\left(\frac{d \sigma}{d \Omega} \right)_{c.m.} =
\frac{1}{W_{\pi N}^2 
\, (8 \pi)^2} \, \frac{| \vec{k}^{*}_{\pi} |}{E_\gamma^{c.m.}} \,
\, \frac{1}{4} \, \sum_{\sigma} \sum_{\sigma'} \sum_{\lambda = \pm 1} 
 {\arrowvert \, e \, \varepsilon_\nu(q, \lambda) \, 
<\pi N|j^{\nu}(0)|N> \arrowvert }^2,
\end{eqnarray}
where $E_\gamma^{c.m.}$ is the photon c.m. energy, 
$\varepsilon_\nu(q, \lambda)$ is the photon polarization vector,
$W_{\pi N}$ is the $\pi N$ c.m. energy, 
and $j^\nu$ is the electromagnetic current operator.
We show the results of the different model contributions discussed
above to the pion photoproduction total cross sections
in Fig.~\ref{fig:tot_gapi}.
As can be noticed from Fig.~\ref{fig:tot_gapi}, our estimates 
consisting of a soft-pion production amplitude supplemented by   
a $\Delta$-resonance production mechanism  
reproduce the cross sections on the lower
energy side of the $\Delta(1232)$ resonance. Around resonance
position our simple model overestimates the cross sections by about 10 \%,
which is mainly due to rescattering contributions which can be
included by a proper unitarization of the amplitude, which we did not
perform in our simple estimate. At the higher energy side of the 
$\Delta(1232)$ resonance, the present model somewhat overestimates the
data. Besides the unitarization, this due to the
increasingly important role from the t-channel exchanges 
of $\rho$ and $\omega$ vector mesons in the non-resonant part of the
amplitude. It is known that these vector meson exchanges yield a
destructive interference on the higher energy side of the
$\Delta(1232)$ resonance. 
Since our objective here is not to present a phenomenological model for 
pion photoproduction, able to precisely describe the available data,
but to provide an estimate for the associated DVCS process, we do not
include a proper unitarization or vector meson exchanges as one is not 
able at this point to model the corresponding 
contributions for the ADVCS process. 
Therefore the quality of the description shown in
Fig.~\ref{fig:tot_gapi} is indicative of the quality of our
corresponding estimates for the ADVCS process which are performed
along the same lines, i.e. by the sum of a non-resonant soft-pion production 
amplitude and a $\Delta(1232)$ resonance production amplitude as outlined
above. 
\newline
\indent
In the following calculations we use, for the nucleon GPDs,  the model
described in Refs.~\cite{Vdh99,Kiv01,GPV01}, to which we refer the
reader for details. We construct the GPDs from a double distribution
based on the forward unpolarized quark distributions of
MRST01~\cite{Mar01} (for the GPD $H$) and based on the forward polarized
quark distributions of Ref.~\cite{Lea98} (for the GPD $\tilde H$). 
To construct the double distribution, we use a profile function with
parameters $b_{val} = b_{sea} = 1$ as detailed in Ref.~\cite{GPV01}.
For the $t$-dependence of the GPD (at the moderately small values of
$t$ considered in this paper) we adopt, unless stated otherwise, 
a factorized ansatz by multiplying the $x$ and $\xi$ dependent function by 
the corresponding form factor (in $t$) so as to satisfy the first sum
rule. In the calculations for the beam charge asymmetry, we also
compare the results with a model for the GPDs where a D-term is added
to the double distribution, with the parametrization given 
in Ref.~\cite{Kiv01}.  
For all other calculations, where it is not stated explicitely, the
results do not include a D-term in the model for the GPDs.

\begin{figure}[ht]
{\centering \resizebox*{0.65\textwidth}{!}{\includegraphics{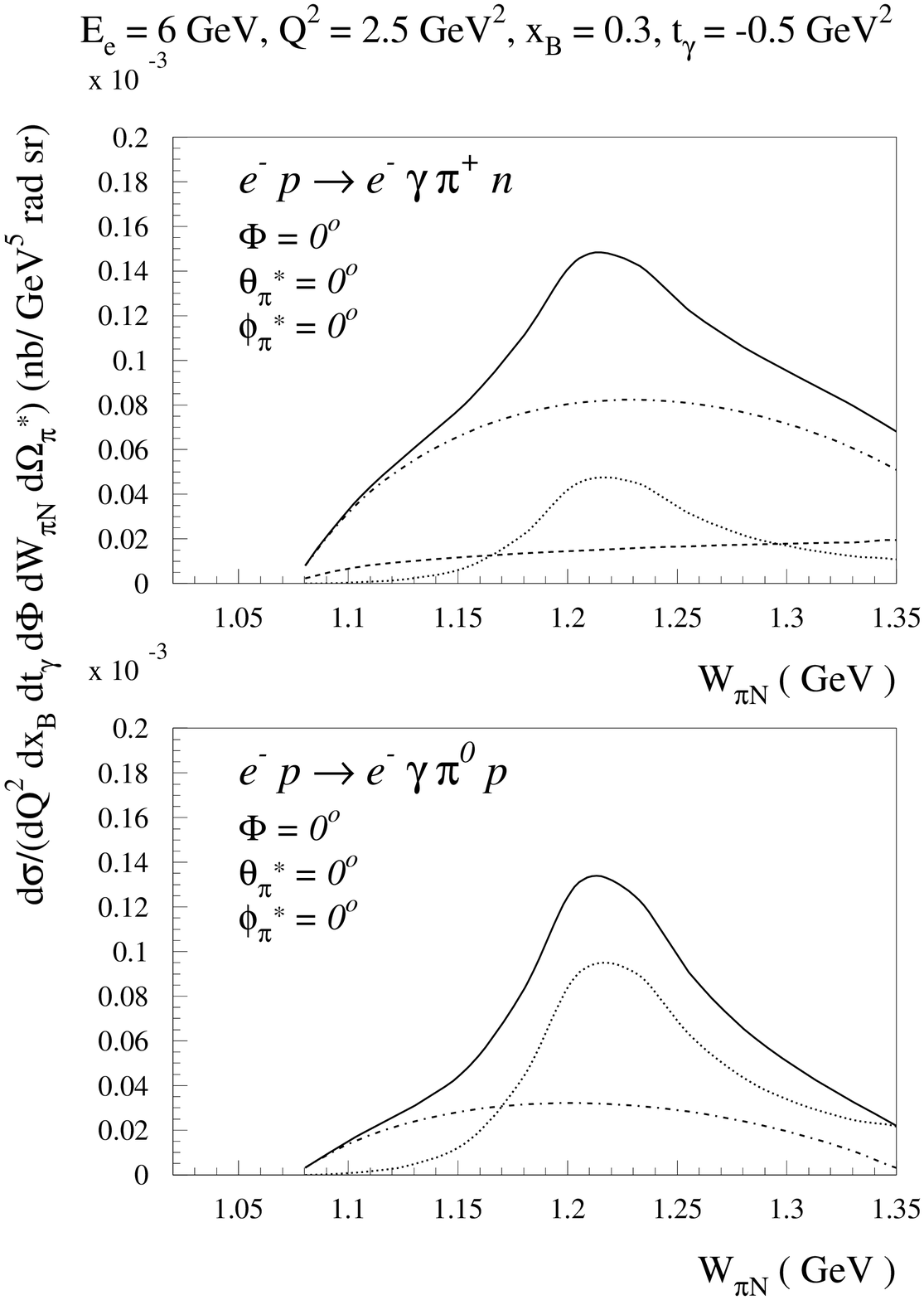}} \par}
\caption{Different contributions to the 
7-fold differential cross sections for the
ADVCS processes in JLab kinematics,  
with pion emitted in the same direction as the recoiling $\pi N$ system 
(corresponding with $\theta_\pi^* = 0^o$). 
Dashed-curve : commutator contribution. 
Dashed-dotted curves : commutator + Born contributions. 
Dotted curves : $\Delta$ contribution. 
Solid curves : commutator + Born + $\Delta$ contributions. 
For the $N \to \Delta$ GPDs both $C_1$ and $H_M$ are included.} 
\label{fig:advcs7f_6a}
\end{figure}

\indent
In Fig.~\ref{fig:advcs7f_6a}, we study the 
different ADVCS processes and show their contributions to the 7-fold
differential $e^-  p \to e^- \gamma \pi N$ cross sections,
differential in $Q^2, x_B, t_\gamma, \Phi$, the $\pi N$ invariant mass 
$W_{\pi N}$, and the pion solid angle $\Omega^*_{\pi}$ in the 
$\pi N$ rest frame.  By comparing Figs.~\ref{fig:tot_gapi} and
\ref{fig:advcs7f_6a}, one sees that the ratio of the non-resonant to resonant
contributions is larger for the ADVCS process compared to the pion
photoproduction process.  

\begin{figure}[ht]
{\centering \resizebox*{0.7\textwidth}{!}{\includegraphics{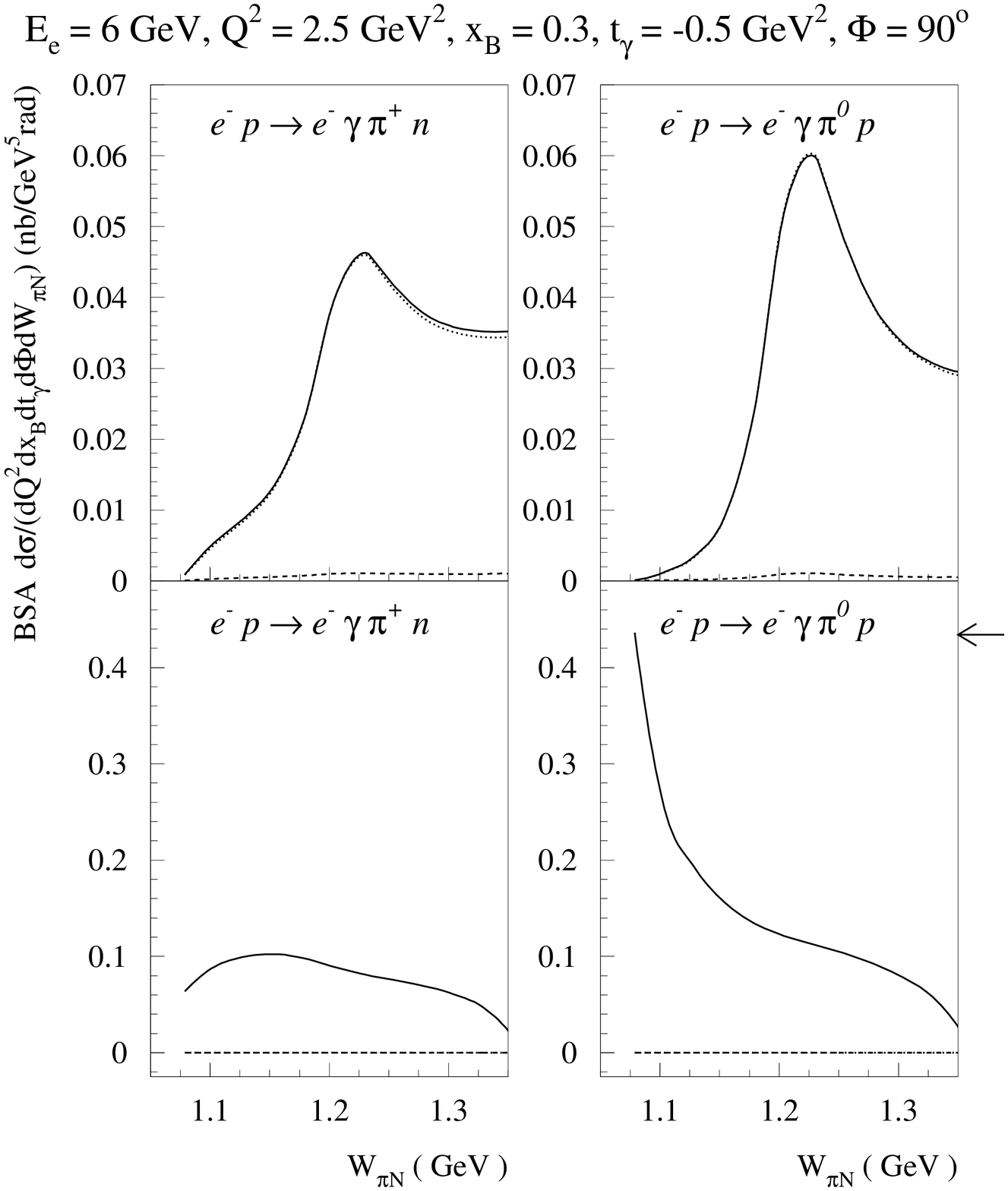}} \par}
\caption{5-fold differential cross sections (upper panels) and
corresponding beam-spin asymmetries (lower panels) for the
$e^- p \to e^- \gamma \pi N$ reactions in JLab kinematics. 
Dotted curves : ABH; dashed curves : ADVCS; solid curves : ABH +
ADVCS.
The arrow gives the elastic value of the BSA for the
BH + DVCS process, corresponding with $W_{\pi N} = M = 0.939$~GeV.}
\label{fig:ssa5f_6}
\end{figure}

\indent
In Fig.~\ref{fig:ssa5f_6}, we compare the 
5-fold differential $e^- p \to e^- \gamma \pi N$
cross sections, 
i.e. integrated over the pion solid angle $\Omega^*_{\pi}$ ,
for the ABH, ADVCS and ABH + ADVCS processes 
\footnote{To simplify the notations, we note ABH + ADVCS for the cross
sections of the coherent sum of both processes.}
for JLab kinematics. Clearly, the ABH largely dominates the
cross sections. The resulting beam spin asymmetries (BSA), for a
polarized lepton beam, are around 5 - 10 \% for the 
$e^- p \to e^- \gamma \pi^+ n$ process. For the 
$e^- p \to e^- \gamma \pi^0 p$ process, the BSA grows 
when approaching the $\pi N$ threshold, where it reaches the same
value as for the $e^- p \to e^- \gamma p$ process.
This can be easily understood because at threshold, the amplitude for
the $e p \to e \gamma \pi^0 p$ process is obtained 
from the $e p \to e \gamma p$ process by attaching a
soft pion to the initial and final proton. 
This is what we called the Born term in Section 5
(Fig.~\ref{Fig_soft_pion} a). This amounts to multiply
the DVCS and BH amplitudes of the $e p \to e \gamma p$ process by
the same factor when calculating their $e p \to e \gamma \pi^0 p$
counterparts. Therefore, when taking the ratio of cross sections in
the BSA, which is due to the interference of ABH and ADVCS, this common
factor drops out and one obtains the same BSA as for the $e^- p \to
e^- \gamma p$ process. Note that this is not the case for the $e^- p
\to e^- \gamma \pi^+ n$ process, where both
commutator and Born terms contribute. Futhermore in the Born
term  for the $p \to \pi^+ n$ transition, amplitudes involving both
proton and neutron GPDs interfere according to whether the charged
pion is emitted from the final or intial nucleons respectively. This
results in a much smaller BSA at threshold 
for charged as compared to neutral pion production. 
When moving to higher values of $W_{\pi N}$ the $\Delta$ contribution
becomes important, and the ratio of ADVCS to ABH changes compared to
the  $e^- p \to e^- \gamma p$ process. Around $\Delta(1232)$ resonance
position, the BSA for the  charged and neutral pion production
channels reach comparable values, around 10 \%.

\begin{figure}[ht]
{\centering \resizebox*{0.7\textwidth}{!}{\includegraphics{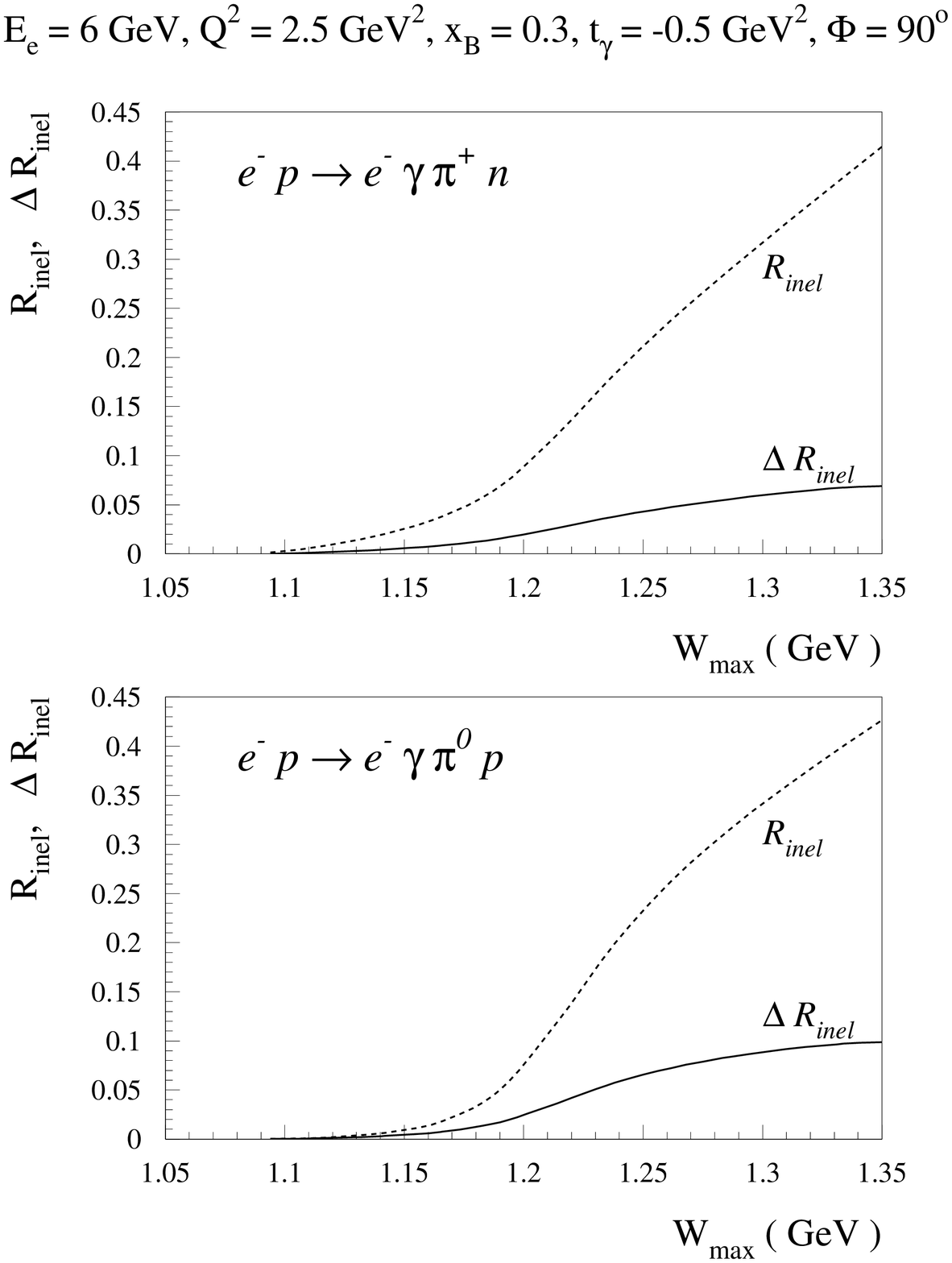}} \par}
\caption{Ratios $R_{inel}$ and $\Delta R_{inel}$  of 
$e^- p \to e^- \gamma \pi N$ (ABH + ADVCS) to $e^- p \to e^- \gamma p$ (BH + DVCS) cross sections 
according to Eqs.~(\ref{eq:R_inel}, \ref{eq:delR_inel}) 
as function of the upper integration limit $W_{max}$ in JLab
kinematics. Dashed (solid) curves : ratio of unpolarized (polarized)
cross sections respectively.}
\label{fig:ssa4f_6}
\end{figure}

\indent
If one does not perform a fully exclusive DVCS experiment, 
one actually measures the cross section 
\begin{eqnarray}
\frac{d \sigma^{exp}}{dQ^2 \, dx_B \, dt \, d\Phi} \,=\,
\frac{d \sigma (e p \to e \gamma p)}{dQ^2 \, dx_B \, dt \, d\Phi}
\,\left( 1 \,+\, R_{inel} \right), 
\label{eq:cs_exp}
\end{eqnarray}
with $R_{inel}$ the ratio of the integrated inelastic $e p \to e \gamma \pi N$
cross section to the cross section for the 
$e p \to e \gamma p$ reaction (i.e. the `elastic' DVCS process), given
by 
\begin{eqnarray}
R_{inel} \,=\, \left( \int_{M + m_\pi}^{W_{max}} \, d W_{\pi N} \, 
\frac{d \sigma (e p \to e \gamma \pi N)}
{dQ^2 \, dx_B \, dt \, d\Phi \, d W_{\pi N}} \right)
\,/\,  
\frac{d \sigma (e p \to e \gamma p) }{dQ^2 \, dx_B \, dt \, d\Phi }\, .
\label{eq:R_inel}
\end{eqnarray}
This  is the ratio $\kappa$ introduced in Eq.~(\ref{Eq_prelim.10}), but written in
a more familiar way. 
The ratio $R_{inel}$ depends upon the upper integration limit $W_{max}$,
determined by the resolution of the experiment. 
\newline
\indent
We can now provide an estimate of the `contamination' to the 
BSA for a not fully exclusive experiment. In an experiment where one
does not separate the $\gamma p$ and $\gamma \pi N$ final states, one
actually measures  
\begin{eqnarray}
(BSA)^{exp} \, = \, \frac{\Delta d \sigma^{el} \, \left( 1 \,+\, \Delta R_{inel}\right) }
{2 \,  d \sigma^{el} \, \left( 1 \,+\, R_{inel}  \right)},
\label{eq:bsa_exp}
\end{eqnarray}
where $\Delta d \sigma^{el}$ and $d \sigma^{el}$ stand for 
\begin{eqnarray}
\Delta d \sigma^{el} &=&  
\frac{d \sigma_{h = + 1/2}}{dQ^2 \, dx_B \, dt \, d\Phi} (e p \to e \gamma p)  
- 
\frac{d \sigma_{h = - 1/2}}{dQ^2 \, dx_B \, dt \, d\Phi}  (e p \to e \gamma p),
\nonumber\\ 
d \sigma^{el}  &=& \frac{1}{2} 
\left(  \frac{d \sigma_{h = + 1/2}}{dQ^2 \, dx_B \, dt \, d\Phi}   (e p \to e \gamma p)
+ 
 \frac{d \sigma_{h = - 1/2}}{dQ^2 \, dx_B \, dt \, d\Phi}  (e p \to e \gamma p)
\right),
\end{eqnarray}
with $h$ the lepton beam helicity. 
In Eq.~(\ref{eq:bsa_exp}), $R_{inel}$ is given as in Eq.~(\ref{eq:R_inel}), 
and $\Delta R_{inel}$ is the corresponding ratio of inelastic
to elastic DVCS helicity cross sections~:
\begin{eqnarray}
\Delta R_{inel} \,=\, \left( \int_{M + m_\pi}^{W_{max}} \, d W_{\pi N} \, 
\frac{\Delta d \sigma (e p \to e \gamma \pi N)}
{dQ^2 \, dx_B \, dt \, d\Phi \, d W_{\pi N}} \right)
\,/\,  
\frac{\Delta d \sigma (e p \to e \gamma p) }{dQ^2 \, dx_B \, dt \, d\Phi }.
\label{eq:delR_inel}
\end{eqnarray}
From Eq.~(\ref{eq:bsa_exp}), one sees that for a not fully exclusive
experiment the `elastic' beam-spin asymmetry  $(BSA)^{el}$ 
for the $e p \to e \gamma p$ process is related  
to the measured beam-spin asymmetry $(BSA)^{exp}$ through~: 
\begin{eqnarray}
(BSA)^{el} \, \equiv \, 
\frac{\Delta d \sigma^{el} }{2 \; d \sigma^{el} } 
\, = \,
R_{BSA} \, \cdot \, (BSA)^{exp} ,
\label{eq:bsa_exp2}
\end{eqnarray}
where the correction factor $R_{BSA}$ is given by~:
\begin{eqnarray}
R_{BSA} \, = \, \frac
{1 \,+\, R_{inel}}{1 \,+\, \Delta R_{inel}} .
\label{eq:rcorr}
\end{eqnarray}
\newline
\indent
In Fig.~\ref{fig:ssa4f_6}, we show the ratios $R_{inel}$ and 
$\Delta R_{inel}$. One sees that when
integrating up to $W_{max} \simeq 1.35$ GeV, the helicity
cross sections ratio $\Delta R_{inel}$ 
reaches about 10 \% for $\gamma \pi^+ n$ and
$\gamma \pi^0 p$ final states separately. 
On the other hand, the unpolarized cross section 
ratio $R_{inel}$ is much larger and reaches about 40 \% for each channel
separately. This difference originates in the different ratio of the
ABH cross section as compared to the corresponding ADVCS ratio.
This different ratio for the ABH and ADVCS processes has as consequence
that the BSA, which is due to an interference of both, receives an
important correction in an experiment where the final state cannot
be fully resolved. 

\begin{figure}[ht]
{\centering \resizebox*{0.7\textwidth}{!}{\includegraphics{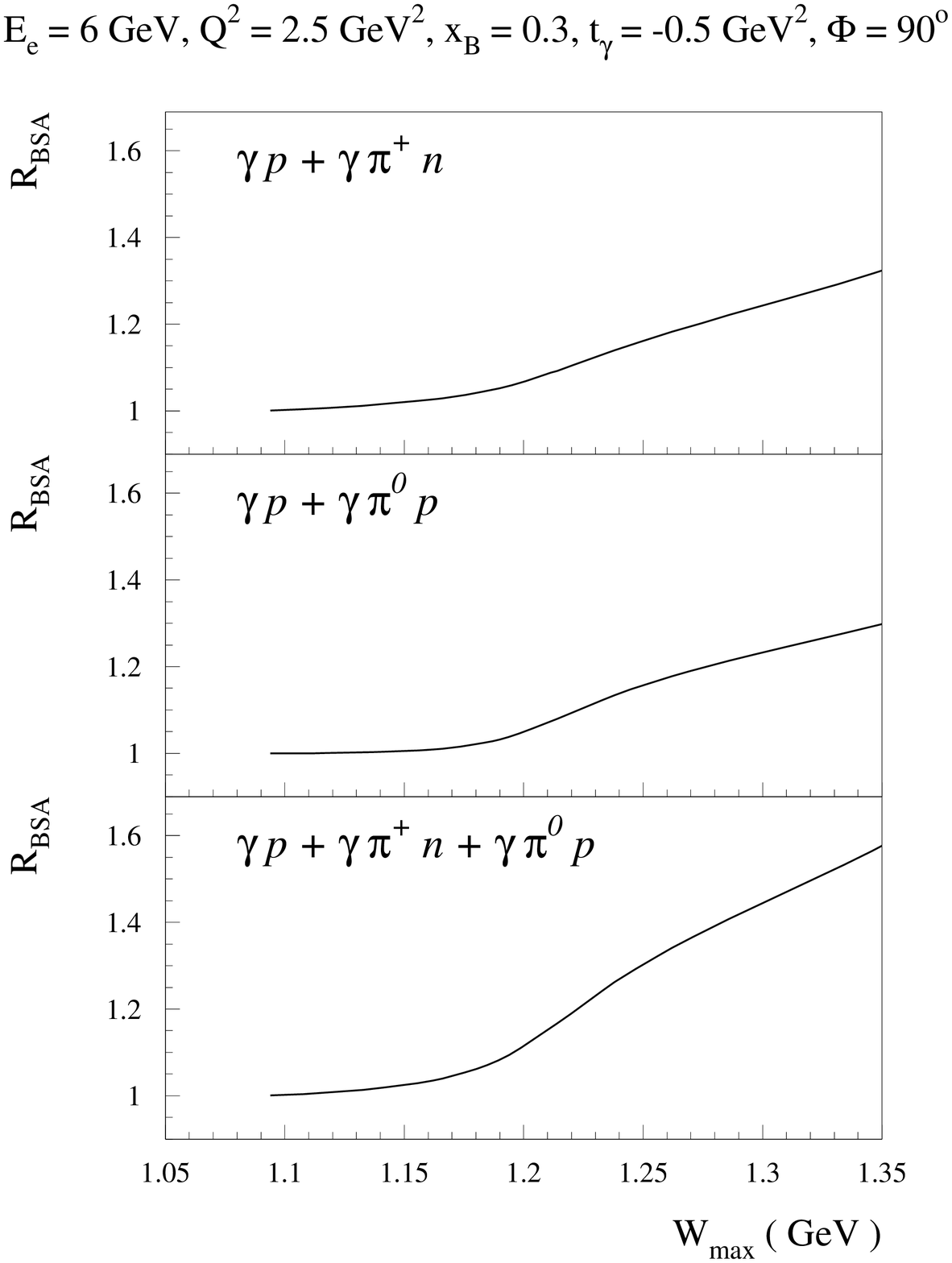}} \par}
\caption{Correction factor according to Eq.~(\ref{eq:rcorr}) 
to apply to the measured BSA 
to extract the $e^- p \to e^- \gamma p$ BSA, 
when the measurement is not fully exclusive. 
The correction factors are given 
for the situations where the $\gamma p$ final state is
contaminated either by $\gamma \pi^+ n$ (upper panel), by 
$\gamma \pi^0 p$ (middle panel), or by both 
$\gamma \pi^+ n$ and $\gamma \pi^0 p$ (lower panel). The correction is
given as function of the upper integration limit $W_{max}$ in JLab
kinematics. }
\label{fig:ssa_corr6}
\end{figure}

\indent
In Fig.~\ref{fig:ssa_corr6}, we show the resulting correction factor
for the BSA defined in Eq.~(\ref{eq:rcorr}) for JLab kinematics. 
For an experiment which measures an $e^-$ and a proton, 
and which reconstructs the final 
$\gamma$ from the missing mass, 
but where the resolution does not permit to fully separate 
the $\gamma p$ final state from a $\gamma \pi^0 p$ final state, 
such as in the bulk of the events of 
the first DVCS experiment at CLAS \cite{Step02}, 
one obtains a correction factor of around 1.3 when 
integrating up to  $W_{max} \simeq 1.35$ GeV 
in the kinematics considered in Fig.~\ref{fig:ssa_corr6}. 
For an experiment which only measures an $e^-$ and a $\gamma$ in the final
state and where the resolution does not permit to separate the $p$ 
hadronic final state from the $\pi^+ n$ and $\pi^0 p$ hadronic final states, the
correction factor, integrated up to  $W_{max} \simeq 1.35$ GeV,
amounts to about 1.6 in the same kinematics.

\begin{figure}[ht]
{\centering \resizebox*{0.7\textwidth}{!}{\includegraphics{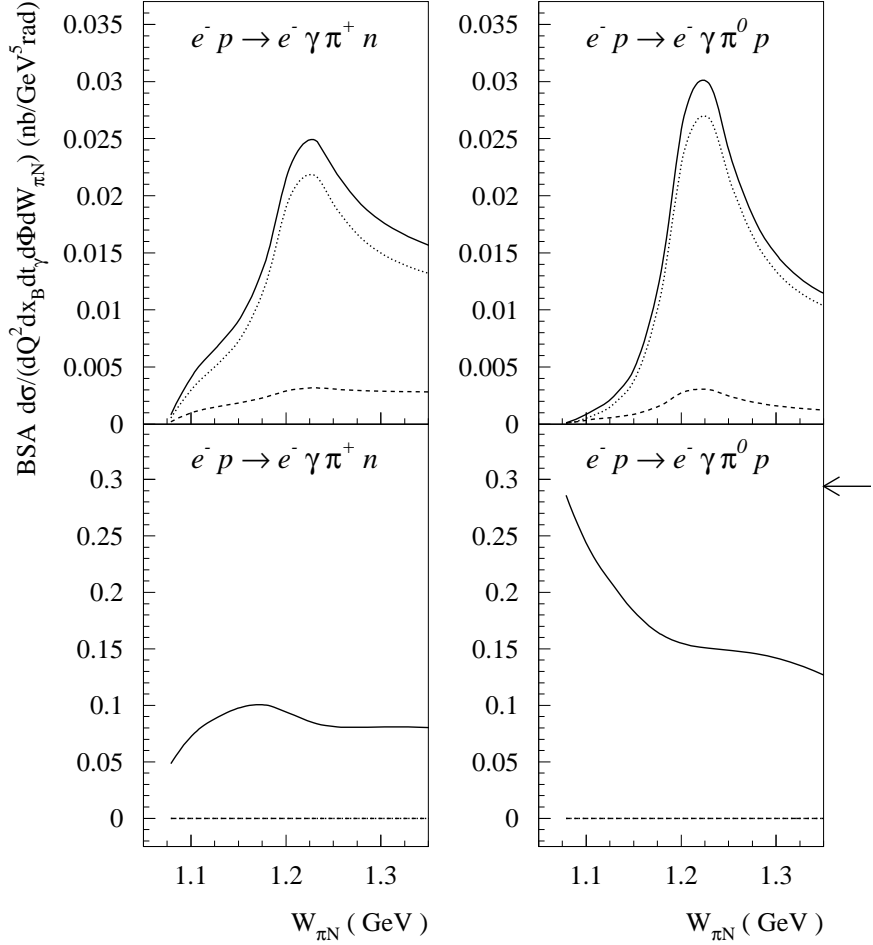}} \par}
\caption{5-fold differential cross sections (upper panels) and
corresponding beam-spin asymmetries (lower panels) for the
$e^- p \to e^- \gamma \pi N$ reactions in HERMES kinematics. 
Dotted curves : ABH; dashed curves : ADVCS; solid curves : ABH + ADVCS.
The arrow gives the elastic value of the BSA for the
BH + DVCS process, corresponding with $W_{\pi N} = M = 0.939$~GeV.}
\label{fig:ssa5f_27}
\end{figure}

\begin{figure}[ht]
{\centering \resizebox*{0.7\textwidth}{!}{\includegraphics{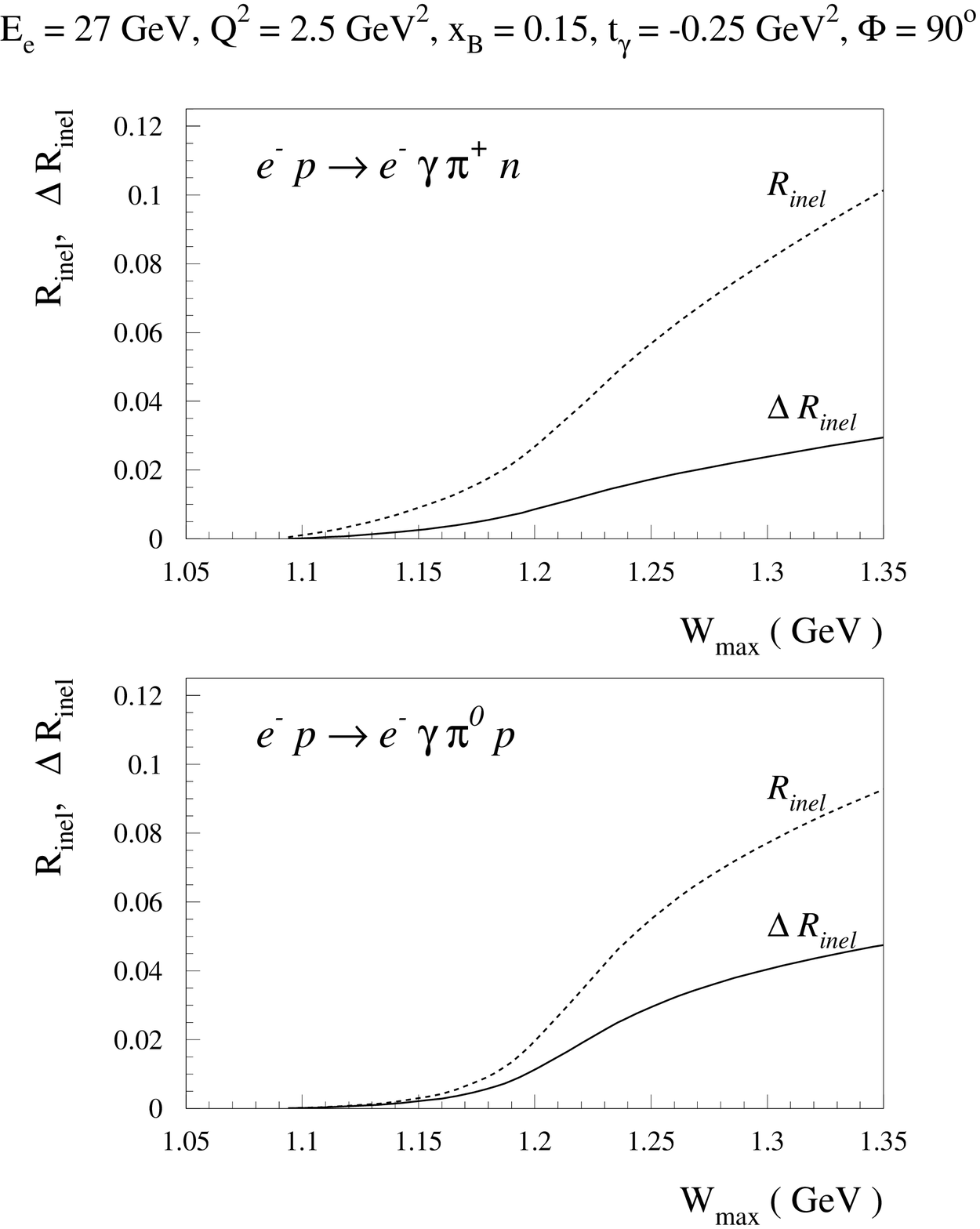}} \par}
\caption{Ratio of $e^- p \to e^- \gamma \pi N$ (ABH + ADVCS) 
to $e^- p \to e^- \gamma p$ (BH + DVCS) cross sections 
according to Eqs.~(\ref{eq:R_inel}, \ref{eq:delR_inel}) 
as function of the upper integration limit $W_{max}$ in HERMES
kinematics. Dashed (solid) curves : ratio of unpolarized (polarized)
cross sections respectively.}
\label{fig:ssa4f_27}
\end{figure}

\begin{figure}[ht]
{\centering \resizebox*{0.7\textwidth}{!}{\includegraphics{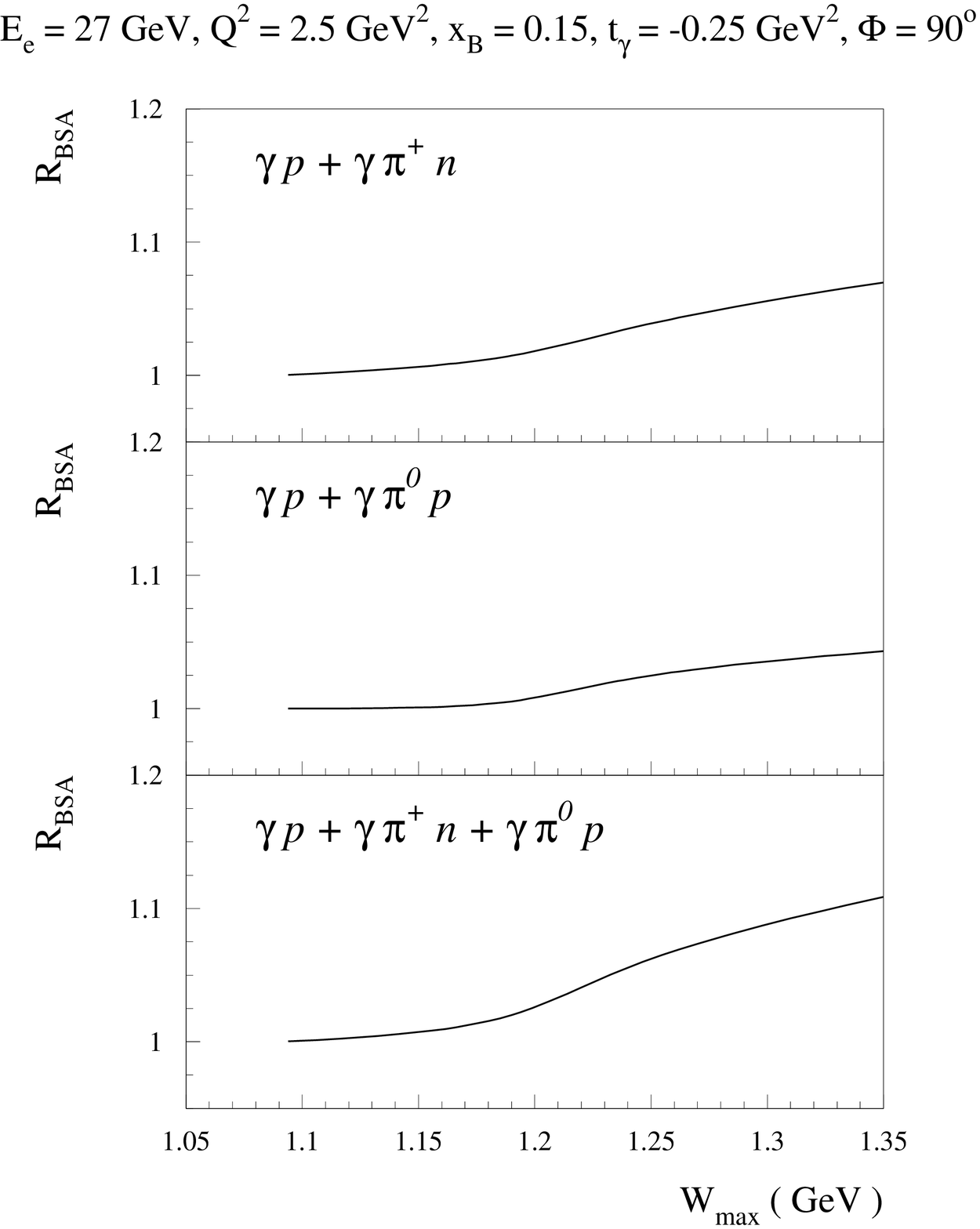}} \par}
\caption{Correction factor according to Eq.~(\ref{eq:rcorr}) 
to apply to the measured BSA 
to extract the $e^- p \to e^- \gamma p$ BSA, 
when the measurement is not fully exclusive. 
The correction factors are given 
for the situations where the $\gamma p$ final state is
contaminated either by $\gamma \pi^+ n$ (upper panel), by 
$\gamma \pi^0 p$ (middle panel), or by both 
$\gamma \pi^+ n$ and $\gamma \pi^0 p$ (lower panel). The correction is
given as function of the upper integration limit $W_{max}$ for HERMES
kinematics. }
\label{fig:ssa_corr27}
\end{figure}

\newpage
In Figs.~\ref{fig:ssa5f_27} - \ref{fig:ssa_corr27}, we study the
corresponding effects in the kinematics accessible in the HERMES experiment.
\newline
\indent
One sees from Fig.~\ref{fig:ssa5f_27} that in the kinematics
accessible at HERMES, 
the ABH still dominates the cross sections. 
The interference of the ABH and ADVCS
processes leads to a BSA which is around 10\% for the 
$e^- p \to e^- \gamma \pi^+ n$ reaction. For the $e^- p \to e^- \gamma 
\pi^0 p$ reaction, the BSA rises towards $\pi N$ threshold where it
reaches the value of the BSA for the $e^- p \to e^- \gamma p$
reaction, as discussed before. Around the $\Delta(1232)$ resonance
position, the BSA for the $e^- p \to e^- \gamma \pi^0 p$ reaction
reaches about 15 \%.
\newline
\indent
For the present DVCS experiments at HERMES \cite{Air02} 
where the experimental resolution does not allow to fully reconstruct the 
final state, it is important to estimate the
contribution of the $\gamma \pi^+ n$ and $\gamma \pi^0 p$ final
states, which we show in Fig.~\ref{fig:ssa4f_27}. 
When integrating the cross sections 
up to $W_{max} \simeq 1.35$ GeV, the helicity
cross sections ratio $\Delta R_{inel}$ 
reaches about 3 \% (5 \%) for the $\gamma \pi^+ n$ ($\gamma \pi^0 p$) 
final states respectively. 
On the other hand, the unpolarized cross section 
ratio $R_{inel}$ reaches about 10 \% for each channel
separately. This different ratio leads to a correction for the 
BSA in an experiment where the final state cannot
be fully resolved, which is shown in Fig.~\ref{fig:ssa_corr27}.  
For kinematics close to the first DVCS experiment at HERMES \cite{Air02}  
which measured an $e^-$ and a $\gamma$ 
and where the resolution did not permit to separate the $p$ 
hadronic final state from the $\pi^+ n$ and $\pi^0 p$ hadronic final states,
one sees that the correction factor on the BSA due to the $\Delta$
resonance region, i.e.  integrated up to  $W_{max} \simeq 1.35$ GeV, 
amounts to about 1.1 .

\begin{figure}[ht]
{\centering \resizebox*{0.7\textwidth}{!}{\includegraphics{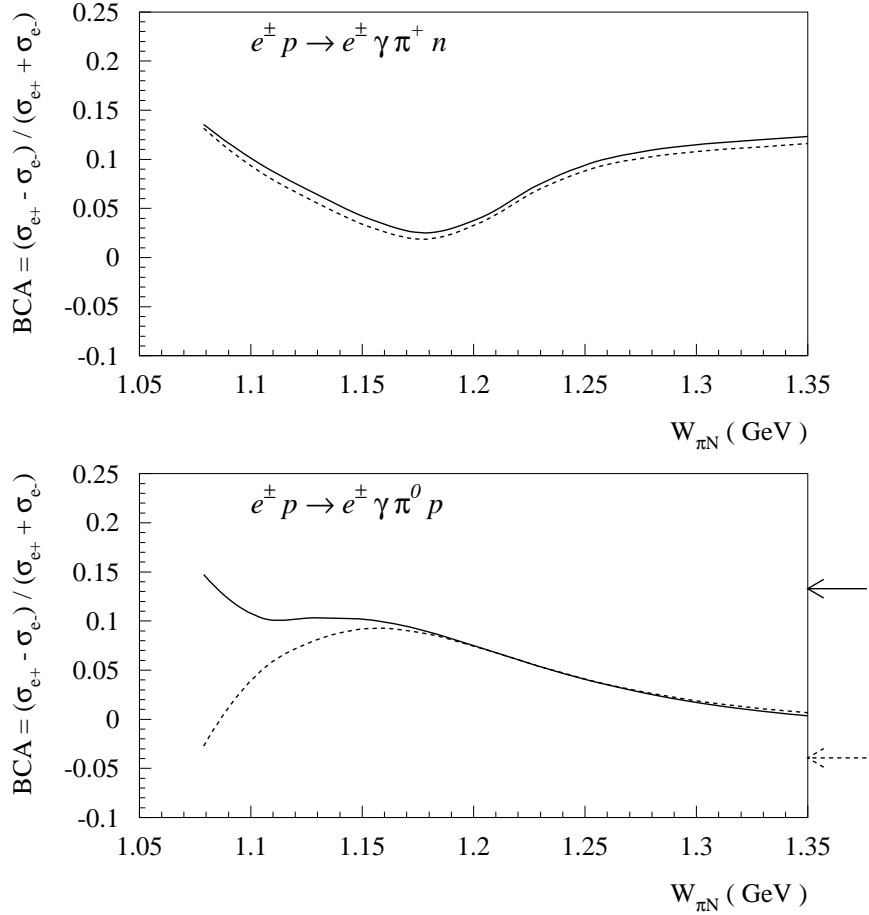}} \par}
\caption{Beam charge asymmetries (BCA) for the
ABH + ADVCS processes in HERMES kinematics. 
The result is shown for 2 different models of the GPDs : without
D-term (dashed curves) and with D-term (solid curves). 
The corresponding arrows give the elastic value of the BCA for the
BH + DVCS process, corresponding with $W_{\pi N} = M = 0.939$~GeV.}
\label{fig:bca5f_27}
\end{figure}

\begin{figure}[ht]
{\centering \resizebox*{0.7\textwidth}{!}{\includegraphics{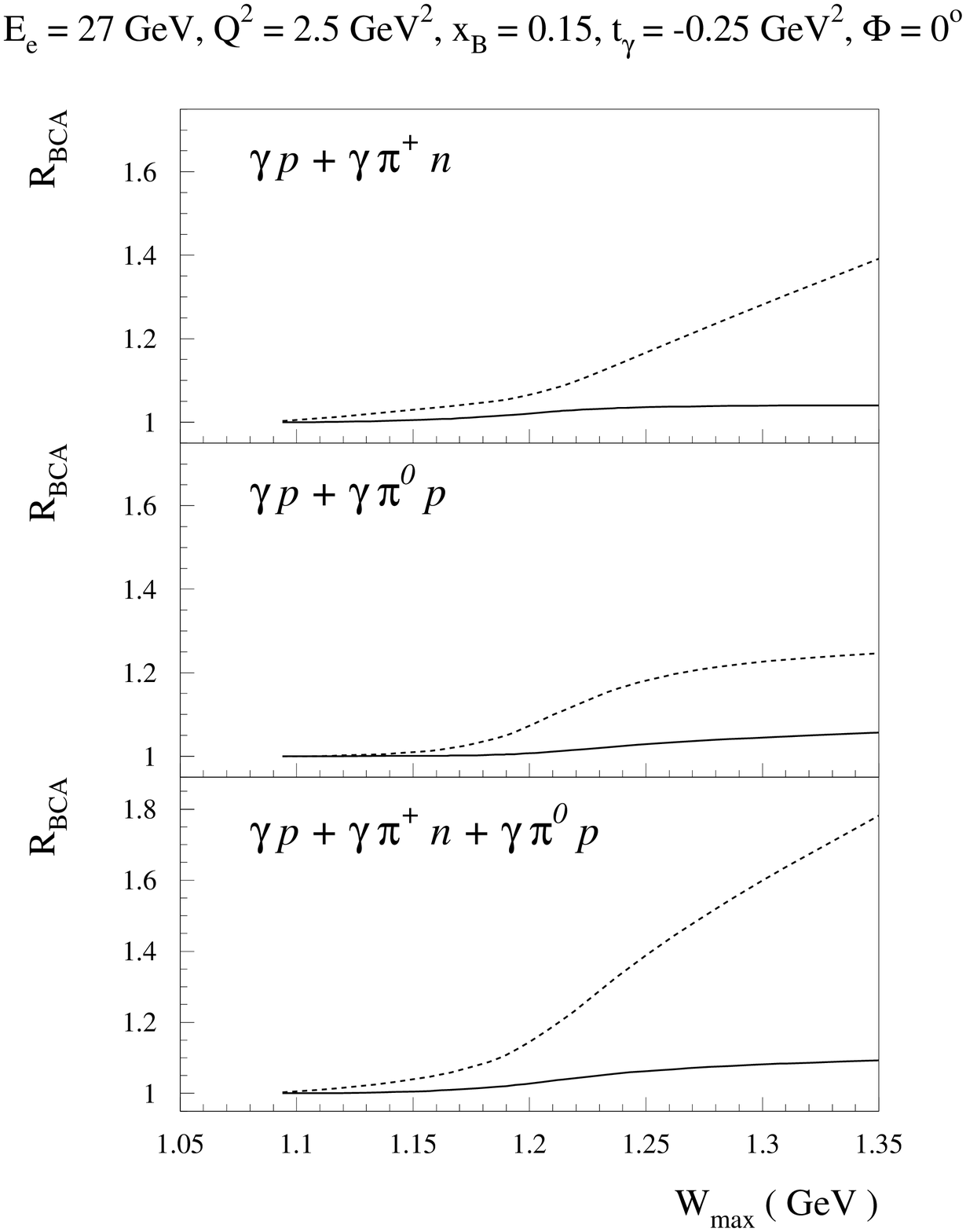}} \par}
\caption{Correction factor according to Eq.~(\ref{eq:bca_rcorr}) 
to apply to the measured BCA 
to extract the BCA between the `elastic' processes $e^\pm p \to e^\pm \gamma p$, 
when the measurement is not fully exclusive. 
The correction factors are given 
for the situations where the $\gamma p$ final state is
contaminated either by $\gamma \pi^+ n$ (upper panel), by 
$\gamma \pi^0 p$ (middle panel), or by both 
$\gamma \pi^+ n$ and $\gamma \pi^0 p$ (lower panel). The correction is
given as function of the upper integration limit $W_{max}$ for HERMES
kinematics. The solid and dashed curves correspond with the 
two GPD models as described in Fig.~\ref{fig:bca5f_27}.}
\label{fig:bca_corr27}
\end{figure}

\indent
We next discuss the beam charge asymmetries (BCA) between the 
$e^+ p \to e^+ \gamma \pi N$  and  $e^- p \to e^- \gamma \pi N$
processes.  The BCA in kinematics accessible at HERMES is shown in
Fig.~\ref{fig:bca5f_27} for two models of the GPDs,
one including the D-term and one without the D-term contribution. 
As for the BSA, one sees that for the
$\gamma \pi^0 p$ final state,  the BCA reaches the same value as for the
elastic DVCS process when approaching the $\pi N$ threshold. 
Furthermore, one sees that since the D-term only
contributes to the Born terms, it mainly manifests itsef in 
the neutral pion production channel around threshold, while its effect
is very small on the charged pion production channel. Around
$\Delta(1232)$ resonance, the effect of the D-term is negligible.  
\newline
\indent 
The BCA between the $e^+ p \to e^+ \gamma X$ and $e^- p \to e^- \gamma
X$ processes has been measured at HERMES~\cite{Ell02}. Because this 
experiment does not allow to distinguish the
hadronic final states $X = p$ from $X = \pi^+ n$ and $X = \pi^0 p$, it is of
interest to estimate the `contamination' by the associated pion
production. An experiment which does not separates
$X = p$ from $X = \pi N$ measures~: 
\begin{eqnarray}
(BCA)^{exp} \, = \, 
\frac{\sigma_{e^+}^{el}  \, \left( 1 \,+\, R_+ \right) \; -\;\sigma_{e^-}^{el}  \, \left( 1 \,+\, R_- \right)}
{\sigma_{e^+}^{el}  \, \left( 1 \,+\, R_+ \right) \; +\;\sigma_{e^-}^{el}  \, \left( 1 \,+\, R_- \right)},
\label{eq:bca_exp}
\end{eqnarray}
where $\sigma_\pm^{el}$ stands for the cross section  
$d \sigma  (e^\pm p \to e^\pm \gamma p) / (dQ^2 \, dx_B \, dt \,
d\Phi)$ of the 'elastic' process, 
and where the ratios $R_\pm$ stand for~:
\begin{eqnarray}
R_\pm \,=\, \frac{1}{\sigma_{e^\pm}^{el}} \; \int_{M + m_\pi}^{W_{max}} \, d W_{\pi N} \, 
\frac{d \sigma (e^\pm p \to e^\pm \gamma \pi N)}
{dQ^2 \, dx_B \, dt \, d\Phi \, d W_{\pi N}} .
\label{eq:R_pm}
\end{eqnarray}
From Eq.~(\ref{eq:bca_exp}), one sees that for a not fully exclusive
experiment the `elastic' $(BCA)^{el}$ is obtained 
from the measured $(BCA)^{exp}$ through~: 
\begin{eqnarray}
(BCA)^{el} \, \equiv \, 
\frac{\sigma_{e^+}^{el}  \; -\;\sigma_{e^-}^{el}}{\sigma_{e^+}^{el} \; +\;\sigma_{e^-}^{el}}
\, = \,
R_{BCA} \, \cdot \, (BCA)^{exp} ,
\label{eq:bca_exp2}
\end{eqnarray}
where the correction factor $R_{BCA}$ is given by~:
\begin{eqnarray}
R_{BCA} \, = \, \frac
{1 \,+\, ( \sigma_{e^+}^{el} R_+ \,+\, \sigma_{e^-}^{el} R_- ) / (\sigma_{e^+}^{el} \,+\, \sigma_{e^-}^{el})}
{1 \,+\, ( \sigma_{e^+}^{el} R_+ \,-\, \sigma_{e^-}^{el} R_- ) / (\sigma_{e^+}^{el} \,-\, \sigma_{e^-}^{el})}
 .
\label{eq:bca_rcorr}
\end{eqnarray}
\indent
The correction factor  $R_{BCA}$ is shown in Fig.~\ref{fig:bca_corr27}
for HERMES kinematics.  
For an experiment which does not separate the
$p$ hadronic final state from the $\pi^+ n$ and $\pi^0 p$ hadronic
final states, the correction factor on the BCA, 
integrated up to  $W_{max} \simeq 1.35$ GeV, 
amounts to about 1.8 for the model without D-term and 
reaches around 1.1 for the model with D-term. 
The much smaller correction in the presence of the D-term can be
understood as in this case the elastic BCA has the same sign and
similar magnitude as the inelastic BCA around $\Delta(1232)$ resonance. 
On the other hand, in the absence of the D-term contribution, the
elastic BCA is small and negative, yielding  
a significant different result from the positive BCA around 
$\Delta(1232)$ resonance. Therefore, 
the resulting correction factor is much larger
for the GPD model without D-term. As the correction of the BCA can be
sizeable according to the model for the GPDs, 
this clearly calls for a fully exclusive measurement to separate the
different final states, in order to reliably extract information on
the GPDs. Such an experiment is planned in the near future at
HERMES using a recoil detector~\cite{Herrec}.

\begin{figure}[ht]
{\centering \resizebox*{0.7\textwidth}{!}{\includegraphics{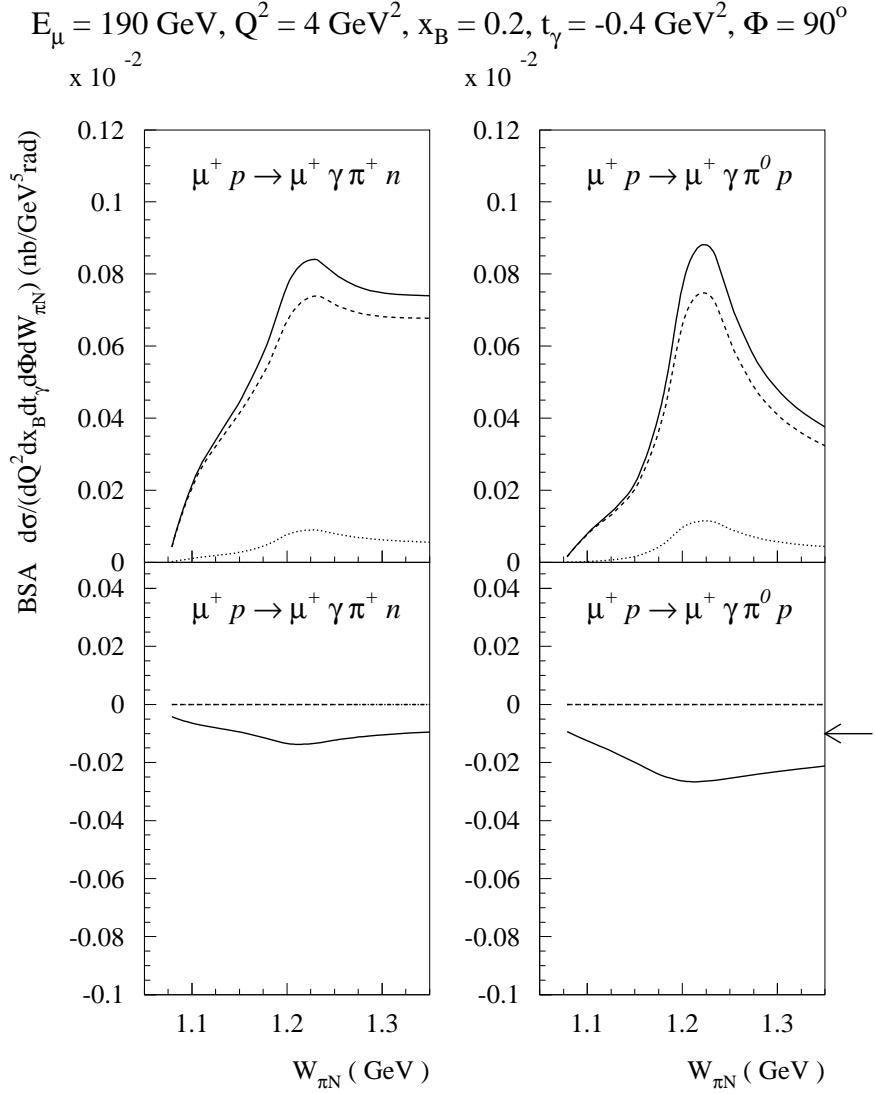}} \par}
\caption{5-fold differential cross sections (upper panels) and
corresponding beam-spin asymmetries (lower panels) for the
$\mu^+ p \to \mu^+ \gamma \pi N$ reactions in COMPASS kinematics. 
Dotted curves : ABH; dashed curves : ADVCS; solid curves : ABH + ADVCS.
The arrow gives the elastic value of the BSA for the
BH + DVCS process, corresponding with $W_{\pi N} = M = 0.939$~GeV.}
\label{fig:ssa5f_190}
\end{figure}

\begin{figure}[ht]
{\centering \resizebox*{0.7\textwidth}{!}{\includegraphics{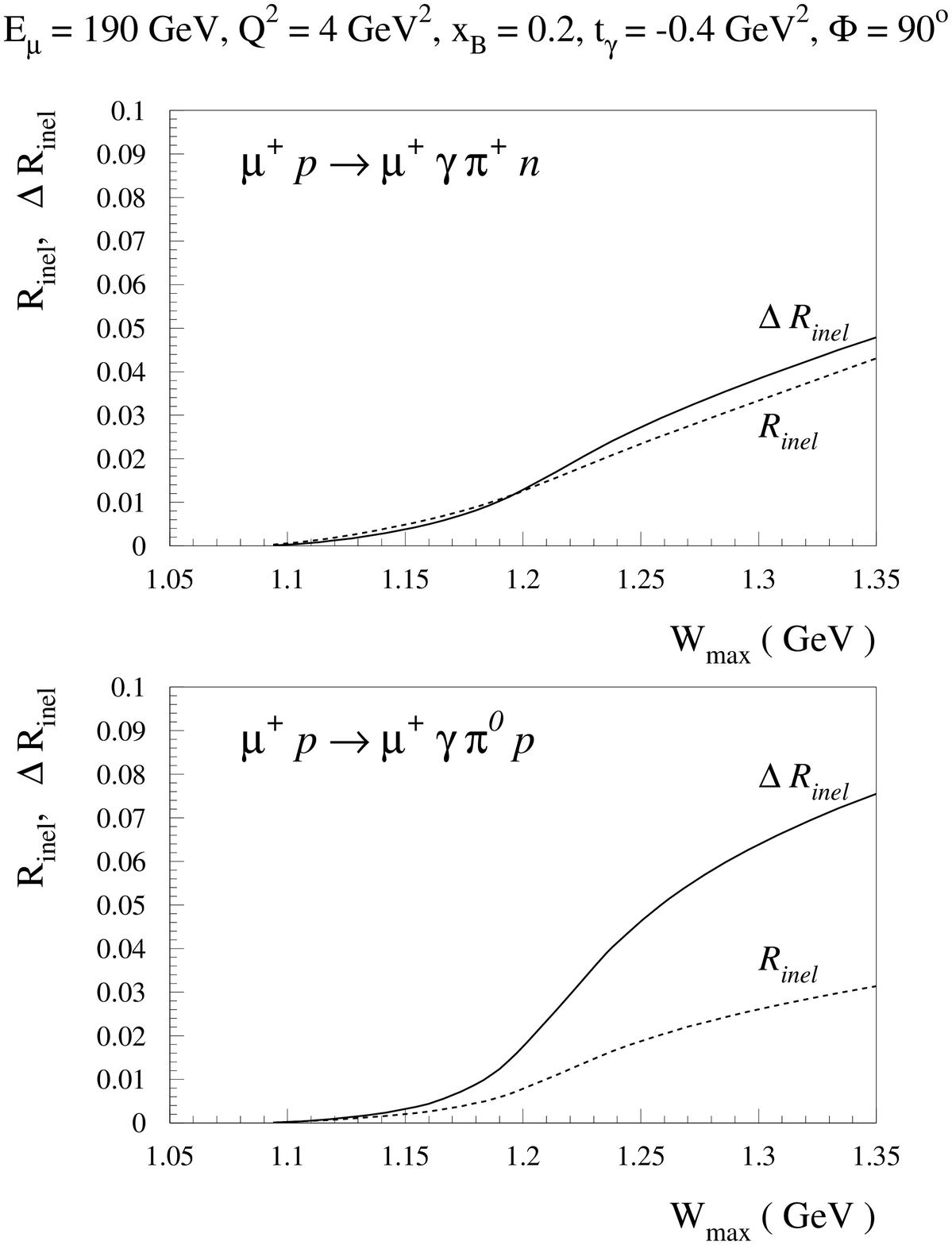}} \par}
\caption{Ratio of $\mu^+ p \to \mu^+ \gamma \pi N$ (ABH + ADVCS) 
to $\mu^+ p \to \mu^+ \gamma p$ (BH + DVCS) cross sections 
according to Eqs.~(\ref{eq:R_inel}, \ref{eq:delR_inel}) 
as function of the upper integration limit $W_{max}$ in COMPASS
kinematics. Dashed (solid) curves : ratio of unpolarized (polarized)
cross sections respectively.}
\label{fig:ssa4f_190}
\end{figure}

\begin{figure}[ht]
{\centering \resizebox*{0.7\textwidth}{!}{\includegraphics{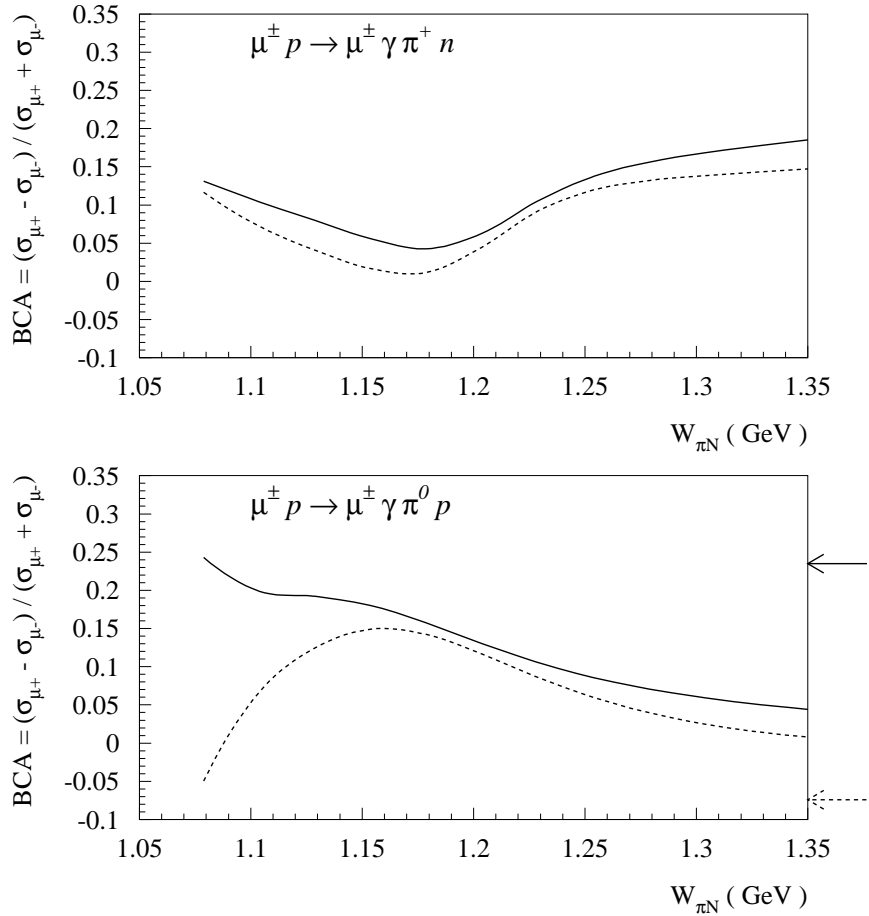}} \par}
\caption{Beam charge asymmetries (BCA) for the
ABH + ADVCS processes in COMPASS kinematics. 
The result is shown for 2 different models of the GPDs. 
Dashed curve : GPD model with factorized t-dependence and without D-term.
Solid curve : GPD model with unfactorized t-dependence and with D-term.
The corresponding arrows give the elastic value of the BCA for the
BH + DVCS process, corresponding with $W_{\pi N} = M = 0.939$~GeV.}
\label{fig:bca5f_100}
\end{figure}

\begin{figure}[ht]
{\centering \resizebox*{0.7\textwidth}{!}{\includegraphics{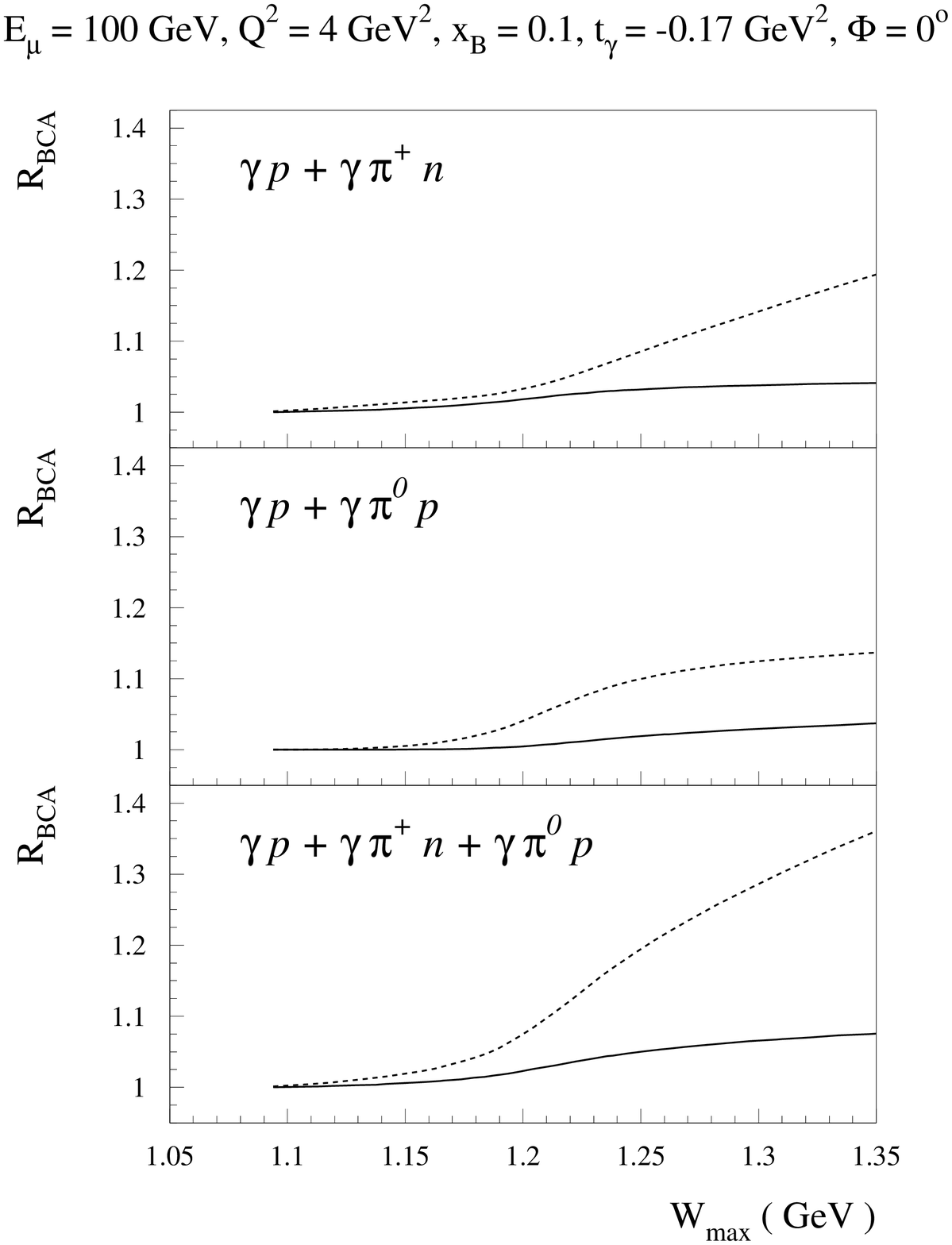}} \par}
\caption{Correction factor according to Eq.~(\ref{eq:bca_rcorr}) 
to apply to the measured BCA 
to extract the BCA between the `elastic' processes $\mu^\pm p \to \mu^\pm \gamma p$, 
when the measurement is not fully exclusive. 
The correction factors are given 
for the situations where the $\gamma p$ final state is
contaminated either by $\gamma \pi^+ n$ (upper panel), by 
$\gamma \pi^0 p$ (middle panel), or by both 
$\gamma \pi^+ n$ and $\gamma \pi^0 p$ (lower panel). The correction is
given as function of the upper integration limit $W_{max}$ for COMPASS
kinematics. The solid and dashed curves correspond with the 
two GPD models as described in Fig.~\ref{fig:bca5f_100}.}
\label{fig:bca_corr100}
\end{figure}

\newpage
\indent
In Figs.~\ref{fig:ssa5f_190}-\ref{fig:bca_corr100}, we show the 
results for kinematics accessible at COMPASS \cite{Compass}.
Fig.~\ref{fig:ssa5f_190} shows the differential cross section 
and BSA for the $\mu^+ p \to \mu^+ \gamma \pi N$ processes. 
In contrast to the previous results for JLab and HERMES kinematics, we
see that at COMPASS the  $\mu^+ p \to \mu^+ \gamma \pi N$ cross
section is dominated by the ADVCS process. The interference 
with the small BH yields only a
small value for the BSA. Due to the dominance of the ADVCS over the
BH, one sees in Fig.~\ref{fig:ssa4f_190} 
an opposite trend for the ratios $R_{inel}$ and $\Delta R_{inel}$
of Eqs.~(\ref{eq:R_inel}, \ref{eq:delR_inel}) in comparison with the
ones shown in Figs.~\ref{fig:ssa4f_6} and \ref{fig:ssa4f_27} for JLab
and HERMES respectively. The larger value of $\Delta R_{inel}$
compared to $R_{inel}$, in particular for $\mu^+ p \to \mu^+ \gamma
\pi^0 p$, leads to a correction factor $R_{BSA}$ of 
Eq.~(\ref{eq:rcorr}) which is slightly smaller than one. 
For an experiment which does not permit to separate the $p$ 
hadronic final state from the $\pi^+ n$ and $\pi^0 p$ hadronic final states,
the correction factor on the BSA due to the $\Delta$
resonance region, i.e.  integrated up to  $W_{max} \simeq 1.35$ GeV, 
amounts to a value around 0.95 at COMPASS.
\newline
\indent
In Figs.~\ref{fig:bca5f_100}, \ref{fig:bca_corr100} we study the 
BCA at COMPASS. 
To obtain sizeable interferences with the BH process, we show the results for 
a lower beam energy of 100 GeV as also accessible at COMPASS. 
We compare the results for two models of the GPDs. 
The first model consists of a
factorized ansatz for the $t$-dependence (compared to the $x$- and
$\xi$ dependences) of the GPDs as used in the
previous calculations, and does not include the D-term. 
The second model includes the D-term and uses an unfactorized Regge
ansatz for the GPDs as specified in Ref.~\cite{GPV01}.
As can be seen from Fig.~\ref{fig:bca5f_100}, 
the $\pi^0$ production process is mainly sensitive to the differences
between those models and at the threshold the BCA for the 
$\mu^\pm p \to \mu^\pm \gamma \pi^0 p$ and $\mu^\pm p \to \mu^\pm
\gamma p$ reactions are the same. 
For an experiment which does not separate the
$p$ hadronic final state from the $\pi^+ n$ and $\pi^0 p$ hadronic
final states, we estimate the correction factors on the BCA in 
Fig.~\ref{fig:bca_corr100}. When integrating the $\pi N$ spectrum up 
to  $W_{max} \simeq 1.35$ GeV, the correction factor on the BCA
asymmetry due to both $\pi^0 p$ and $\pi^+ n$ channels 
amounts to about 1.35 for the model without D-term and 
reaches around 1.05 for the model with D-term.

\newpage

\section{Conclusion\label{Conclusion}}

We have developed a model to calculate the cross section for producing
an extra low energy pion in the photon electro-production reaction.
Our primary goal is to provide a reasonable estimate of the contamination
of the 'elastic' process by this associated reaction when the experimental
data are not fully exclusive. For the various observables which are
generally considered of interest we have defined correction factors
by integrating the associated reaction cross sections up to a given
cutoff.

To build our model we have used the time honored soft pion technique
based on current algebra and chiral symmetry. In the case of the DVCS
reaction, which we always consider in the Bjorken limit, we have assumed
that it was possible to first invoke the factorization theorem and
then to use chiral symmetry to evaluate the matrix elements of the
twist two operators involving one soft pion. Our derivation applies to
the kinematical region $m_\pi^2 << -t << Q^2$ of DVCS type processes,
which corresponds with the kinematical range of experiments considered
at JLab, HERMES and COMPASS.  
The order in which one applies the 
chiral limit ($m_\pi \to 0$) and the Bjorken limit ($Q^2 \to \infty$) 
is a point which certainly 
deserves further attention, see e.g. Ref.~\cite{PPS01} where new
low-energy theorems were derived for the $\gamma^* N \to \pi N$
process at large virtualities  
of the photon ($Q^2 >> \Lambda_{QCD}^3 / m_\pi$).   
In this respect, the chiral perturbation theory approach
developed in Refs.~\cite{Chen01,Arndt02} to study the quark mass dependence
of the parton distribution may be useful. 
For such a systematic chiral perturbation expansion to converge, 
one also sees that the momentum transfer $-t$ should be bound from
above, and be smaller than the chiral symmetry breaking scale of order
$(4 \pi f_\pi)^2$. 

The advantage
of our approach is that the GPDs necessary to calculate the associated
pion production are the same as in the elastic DVCS process. So, in
a relative sense, our estimate is largely independent of the details
of the GPDs.

To extend our estimate to pions of higher energy we have added the
P-wave production assuming it is dominated by the \( \Delta (1232) \)
isobar production. Here, following the approach of Ref.~\cite{Fra00} we have
used the large \( N_{c} \) limit to relate the GPDs of the \( N\rightarrow \Delta  \)
transition to the \( N\rightarrow N  \) ones. Again this minimizes
the model dependence of our results. 

For those experiments which do have the resolution to measure the
$\gamma^* p \to \gamma \pi N$ process, in particular in the
$\Delta(1232)$ resonance region \cite{Michel03}, our calculation
provides a prediction using the large  \( N_{c} \) limit for the 
 \( N\rightarrow \Delta  \) GPDs. The measurement 
of the $\gamma^* p \to \gamma \Delta \to \gamma \pi N$ process
holds the prospect to access information on the quark distributions 
in the $\Delta$ resonance, which are totally unknown at present.  

Generally the DVCS amplitude must be added coherently to the Bethe-Heitler
amplitude. This is also true for the associated reactions and we have
used the same approach to estimate the associated BH and DVCS amplitudes.
For the BH amplitude this amounts to calculate the pion electro-production
amplitude and this gives us a chance to rate the validity of our results
by comparing to the existing data. We find that, for the integrated
cross sections, our estimate are probably valid up to \( W_{max}=1.35~{\rm GeV} \). 

We have performed our calculations for a set of kinematical conditions
which are representative of the present or planned experiments
at JLab, HERMES and COMPASS. In a regime where the ADVCS process
dominates the cross section, such as is the case at COMPASS, 
we find that the pionic contamination (integrated up to \(
W_{max}=1.35~{\rm GeV} \)) never exceeds 10\%.
In a kinematical regime where the ABH process dominates, such as is
typically the case at HERMES and in particular at JLab, the pionic
contamination may become much larger and calls for fully exclusive
experiments. In particular,   
it has an effect on the beam spin and beam charge asymmetries. 
For instance the correction to the BSA due to charged
(neutral) pions in JLab kinematics can reach 30\% each for 
an experiment which is not able to distinguish a $N$ from a $\pi N$
for a cutoff \( W_{max}=1.35~{\rm GeV} \).
The effect on the BSA due to the $\pi^0 p$ and $\pi^+ n$ production 
in HERMES kinematics is of the order of 10\%. 
On the other hand the correction factor on the BCA  
to obtain the `elastic' BCA from an experiment not able to distinguish a $N$
from a $\pi N$ can be as large as a factor 
1.8 at HERMES and 1.35 at COMPASS depending on the model for the GPDs.

To summarize, for a cutoff of \( W_{max}=1.35~{\rm GeV} \) the correction
to the cross sections is moderate but for the BSA and BCA 
our calculations indicate that 
it is wise to consider fully exclusive experiments.

\section*{Acknowledgments}
This work was supported 
by the French Commissariat \`a l'Energie Atomique (CEA),  
by the Deutsche Forschungsgemeinschaft (SFB443),
and in part by the European Commission IHP program
(contract HPRN-CT-2000-00130).
The authors thank N. d'Hose, M. Guidal, X. Ji, M. Polyakov, S. Stratmann,
and L. Tiator for useful discussions.

\end{document}